\documentclass[aps,twocolumn,prb,amsmath,amssymb]{revtex4-1}

\usepackage{soul,color}

\usepackage{verbatim}
\usepackage[varg]{txfonts}
\usepackage{color}
\usepackage[colorlinks=true, citecolor=blue]{hyperref}
\usepackage{amsmath}
\usepackage{graphicx}% Include figure files
\usepackage{bm}% bold math
\usepackage{tikz}
\usetikzlibrary{arrows}
\usetikzlibrary{shapes}

\usepackage{dsfont}
\usepackage{calc}

\makeatletter
\def\textSq#1{%
	\begingroup% make boxes and lengths local
	\setlength{\fboxsep}{0.3ex}% SET ANY DESIRED PADDING HERE
	\setbox1=\hbox{#1}% save the contents
	\setlength{\@tempdima}{\maxof{\wd1}{\ht1+\dp1}}% size of the box
	\setlength{\@tempdimb}{(\@tempdima-\ht1+\dp1)/2}% vertical raise
	\raise-\@tempdimb\hbox{\fbox{\vbox to \@tempdima{%
				\vfil\hbox to \@tempdima{\hfil\copy1\hfil}\vfil}}}%
	\endgroup%
}
\makeatother
\def\XXint#1#2#3{{\setbox0=\hbox{$#1{#2#3}{\int}$}
    \vcenter{\hbox{$#2#3$}}\kern-.5\wd0}}

\def\be{\begin{equation}}
\def\ee{\end{equation}}

\def\bi{\begin{itemize}}
    \def\ei{\end{itemize}}
\def\bn{\begin{enumerate}}
    \def\en{\end{enumerate}}
\def\bea{\begin{eqnarray}}
\def\eea{\end{eqnarray}}
\newcommand{\bpm}{\begin{pmatrix}}
    \newcommand{\epm}{\end{pmatrix}}

\def\ba{\begin{array}}
    \def\ea{\end{array}}
\def\bd{\begin{displaymath}}
\def\ed{\end{displaymath}}

\renewcommand{\imath}{\hspace{1pt}\mathrm{i}\hspace{1pt}}

\renewcommand{\Re}{\mathop{\mathrm{Re}}\nolimits}
\renewcommand{\Im}{\mathop{\mathrm{Im}}\nolimits} 

\begin{document}

\title{Nonlinear response of the  Kitaev honeycomb lattice model in a weak magnetic field}

\author{M. Kazem Negahdari}
\affiliation{Department of Physics, Sharif University of Technology, Tehran 14588-89694, Iran}

\author{Abdollah Langari}
\affiliation{Department of Physics, Sharif University of Technology, Tehran 14588-89694, Iran}
\email{langari@sharif.edu}

\begin{abstract}
	We investigate the nonlinear response of  the  Kitaev honeycomb lattice model in a weak magnetic field using the theory of two-dimensional coherent spectroscopy. We observe that at the isotropic point in the non-Abelian phase of this model, the nonlinear spectrum in the 2D frequency  domain  consists of sharp signals that originate from the flux excitations and Majorana bound states. Signatures of different flux excitations can be clearly observed in this spectrum, such that  one can observe evidences of flux states with 4-adjacent, 2-non-adjacent, and 4-far-separated fluxes, which are not
	visible in linear response spectroscopy such as neutron scattering experiments. Moreover, in the Abelian phase we  perceive that  the  spectrum in the  frequency  domain is composed of streak signals. These signals, as in the nonlinear response of the pure Kitaev model, represent a distinct signature of itinerant  Majorana fermions. However, deep in the Abelian phase whenever a Kitaev exchange coupling is much stronger than the others,  the streak signals are weakened and only single sharp spots are seen in the response, which resembles the dispersionless response of the conventional toric code.

\end{abstract}

\maketitle

\section{ Introduction}\label{Introduct}
In a quantum spin liquid (QSL)\cite{Mila_2000,Balents_2010, Savary_2016,Zhou,Knolle_field_2018, kivleson_QSL_2020,wen_book,moessner_moore_2021,Sadrzadeh2015,Sadrzadeh2016,Sadrzadeh2019}, quantum fluctuations overcome the conventional magnetic orders  even at very low temperature.  Accordingly, the description of QSL state falls beyond the framework of the traditional description of magnetic phases.  Novel concepts such as  emergent gauge fields, fractional spin excitations, and practical potential for the realization of reliable  quantum memories have kept 
the study of QSL as an important topic since the introduction of QSL by  Anderson in 1973\cite{anderson_1973}. 
In the search for QSL phases, the Kitaev honeycomb lattice model (KM)\cite{Kitaev_2006} with the exact QSL ground state has opened a promising route to realize a QSL phase in real materials \cite{Jackeli_Khaliullin2009,Jackeli_Khaliullin2,Plumb_Clancy2014,Winter2016_kitaevm,Trebst2017_kitaevm,Winter_2017_kitaevm,Takagi_Takayama2019,Liu_2021_3d_kitaevm}. According to the Kitaev's  parton construction, where each spin on the lattice is replaced by four Majorana fermions, the bond-dependent Ising interactions of the initial Hamiltonian is turned to the hopping Hamiltonian of Majorana fermions in the presence of emergent $ Z_2 $ gauge fields. Applying a  weak magnetic field on KM, the gapless excitations become gapped
and the system is effectively described by the Kitaev model in the presence of three spins interacting term, which we call the extended Kitaev honeycomb lattice model (EKM) for short that is still exactly solvable\cite{Kitaev_2006}. The applied magnetic field enhances the phase diagram of EKM to show both Abelian and non-Abelian gapped phases.
In the non-Abelian phase, the presence of any $2n$ gauge fluxes imposes $n$ Majorana bound states within the gap, which are fingerprints  of the non-Abelian anyons in this model\cite{Kitaev_2006,lahtinen2008}.

The signature of fractional excitations in the Kitaev QSL state has been exhibited with a  broad continuum observed by the 
conventional dynamical probes \cite{Baskaran2007,Knolle_Chalker_Moessner2014,Knolle_Chern_2014,Knolle_Chalker_Moessner2015,Nasu_Udagawa_Vaporization_2014,Perreault_Brent_Knolle_2015,Halasz_Gabor_2016,Perreault_Knolle_2016}. Merely the the observation of continuum spectrum in the  QSL candidate materials \cite{Sandilands_Luke_2015,Banerjee_Bridges_Nature_2016,Glamazda_Lemmens_2016,Banerjee_Jiaqiang_Science_2017,Wang_Zhe_2017,Banerjee_Lampen_npj_2018,Wulferding_Choi_2020,Ruiz_Breznay_2021,Yang_Wang_2022} 
does not determine without ambiguity whether the continuum is due to fractionalized excitations  or damping of usual quasiparticles or other line-width broadening mechanisms. Duo to the  presence of strong geometrical frustration or competing interactions in such materials, the broad continuum response may have a completely different origin from  the physics of quantum spin liquids\cite{winter2017}. 
Hence, introducing new probes and approaches to extract further information and getting clear identifications
of fractional excitations is of utmost importance in this area of research.
In this respect, the study of nonlinear responses using the  two-dimensional coherent spectroscopy (2DCS)\cite{Mukamel_book_1995,Mukamel_Review2000,Jonas_Review_2DCS,Cho_2DCS_2008,Hamm_book_2011,Woerner_2013,Cundiff_Mukamel2013,Lu_Li_2DCS_2017_spin_wave,Cho_book_2019} can provide clear signatures  of fractional excitations\cite{Armitage_nonlin}. Recent study on the 2DCS  of Kitaev model shows distinct signatures of matter Majorana fermions and gauge field  excitations in the form of diagonal streak signals and their intercepts in 2D frequency domain, respectively\cite{choi1}. According to this technique, one can  reveal distinguishable spectroscopic characteristics of different types of gapped spin liquids\cite{Nandkishore_Choi2021}, signatures of interactions in many-body quantum systems\cite{Li_Oshikawa_Lensing2021,Phuc_Trung_Direct2021,Fava_Biswas2021,marginal_Fermi,Oliver_Hart,Fava_Gopalakrishnan} and extract different relaxation times in quantum systems with quenched disorders\cite{Parameswaran_non1}.  Very recently the 2DCS of one-dimensional Ising model has been investigated by implementing matrix-product state numerical simulations \cite{Sim_GiBaik,Gao_Qi}. Nonlinear responses of KM in the context of high-harmonic generation (HHG) has been studied theoretically\cite{Kanega_lin_nonlin} and  anomalous behavior of the Kitaev spin liquid candidate $ \alpha\text{-RuC}\text{l}_3 $  for static nonlinear susceptibilities has also been reported\cite{Holleis_Anomalous_susceptibility2021}. It has been shown that  the anyonic statistic of quasiparticles can be revealed by nonlinear pump-probe spectroscopy\cite{McGinley_fractional_statistics}.  Moreover within nonlinear responses,  the system is driven to higher excited states, so more states are involved and one can extract further  information about  the system.

In this article we investigate the extended Kitaev model to shed more light on its low energy properties as a QSL and answer few questions like: What are the differences between KM and EKM in terms of nonlinear response of 2DCS? and how the anyons and flux excitations show up in the nonlinear spectrum of EKM both in its Abelian and non-Abelian phases? 
In this respect, we explain the nonlinear magnetic susceptibility in terms of 2DCS in Sec.\ref{tdnls}.   
Moreover, to explicitly determine the physical and unphysical
states in the Kitaev's parton construction, we  consider the labelling of spins and periodic boundary conditions introduced in Refs.~[\onlinecite{Pedrocchi_Loss}] and find the relevant physical states of EKM in Sec. \ref{Model}.
We present our numerical results in Sec.\ref{results}, where  the driving pulses and  the recorded magnetization have the same polarization. In the latter section, we explain that 
the diagonal and off-diagonal streak signals, which exist in the response of the pure KM at the isotropic point\cite{choi1} are no longer dominant in the EKM (where they appear with a tiny strength), instead there exist sharp signals due to the in-gap bound states, which have large contribution to the response. We also  observe signatures of different configurations of the flux states -- such that some flux states with
 4-adjacent, 2-non-adjacent, and 4-far-separated fluxes which are not visible in the linear response -- manifest their contribution in the nonlinear response. 
These are new signatures  of the non-Abelian anyons that can be detected by 2DCS technique. 
Moreover, we inspect the cause of two different response in the Abelian phase of EKM.
We conclude and discuss about our findings in Sec.\ref{conclusion}. 

%%%%%%%%%%%%%%%%%%%%%%%%%%%%%%% 
\begin{figure}[t]
	\center
	\includegraphics[scale=0.672]{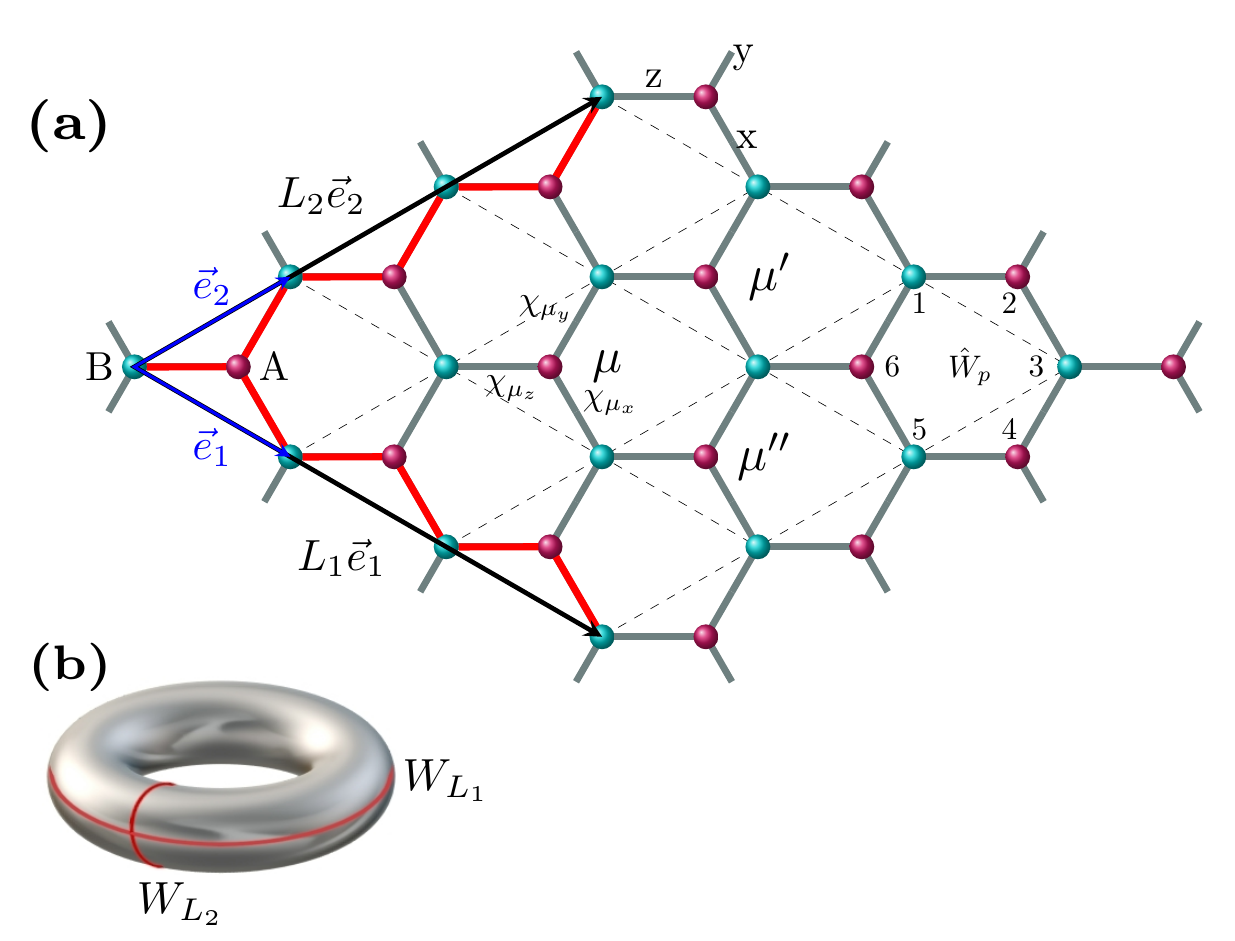}
	\caption{(a) The honeycomb lattice with $L_1=L_2=3 $ and $ M=0 $ according to the explanations in the main text, where $ \mu\ (\mu',\mu'')$ labels the unit cells (dashed
		parallelogram).  The red links in direction $ L_i\boldsymbol{e}_i$ denote the successive terms $ \hat{\sigma}_{i}^{\alpha_{\langle ij\rangle}}\hat{\sigma}_j^{\alpha_{\langle ij\rangle}} $ in the topological loop operator $ \hat{W}_{L_i} $ which also is sketched in (b) for clear observation.}
	\label{lattice}
\end{figure}
%%%%%%%%%%%%%%%%%%%%%%%%%%%%%%

\section{two-dimensional nonlinear spectroscopy}\label{tdnls}

The definition and mathematical expression of the non-linear magnetic susceptibility is given in this section.
The sample is imposed to 
two linearly polarized magnetic impulses  in directions $ \hat{e}_{\alpha} $ and $ \hat{e}_{\beta} $ with $ \tau_1 $ time delay, i.e. $ \boldsymbol{B}(t)=B_{\alpha}\delta(t)\hat{e}_{\alpha}+B_{\beta}\delta(t-\tau_1)\hat{e}_{\beta} $. 
After the second pulse we wait a time interval $\tau_2$ when the magnetization in $\gamma$ direction is measured,
$ M_{\alpha\beta}^{\gamma}(\tau_1+\tau_2) $. In order to eliminate the linear contributions in the response, two separate experiments are performed with a single pulse $ B_{\alpha} $ or $ B_{\beta} $ to measure the  magnetizations $ M_{\alpha}^{\gamma}(\tau_1+\tau_2) $ or $ M_{\beta}^{\gamma}(\tau_1+\tau_2) $. Finally, the nonlinear response, $ M^{\gamma}_{NL}$,
is extracted by removing the linear magnetizations: 
\begin{equation}
 M^{\gamma}_{NL}(\tau_1+\tau_2)=M_{\alpha\beta}^{\gamma}(\tau_1+\tau_2)-M_{\alpha}^{\gamma}(\tau_1+\tau_2)-M_{\beta}^{\gamma}(\tau_1+\tau_2). 
 \label{M_NL}
 \end{equation}
We assume that the system is prepared in its ground state and the magnetization of ground state  is zero, which is the case for QSLs. 
The weak magnetic field is coupled to the magnetization, $ \hat{H}_p=-\boldsymbol{B}(t)\cdot\hat{M}$, 
which is considered in the framework of perturbation theory\cite{choi1}.
Hence, the nonlinear magnetization is obtained as follows:
\begin{align}
\label{non_lin_formula}
M^{\gamma}_{NL}(\tau_1+\tau_2)/2N&=B_{\alpha}B_{\beta}\chi^{(2),\gamma}_{\beta\alpha}(\tau_2,\tau_1)\nonumber\\
&+B_{\beta}B_{\alpha}B_{\alpha}\chi^{(3),\gamma}_{\beta\alpha\alpha}(\tau_2,\tau_1,0)\nonumber\\
&+B_{\beta}B_{\beta}B_{\alpha}\chi^{(3),\gamma}_{\beta\beta\alpha}(\tau_2,0,\tau_1)+\mathcal{O}(B^4),
\end{align}
where N is number of unit cells. The $n$-th order susceptibility  is obtained from the $(n+1)$-points correlation functions,
\begin{equation}
\chi^{(2),\gamma}_{\beta\alpha}(\tau_2,\tau_1)=\frac{-1}{N}\sum_{l=1}^{2}\Re\bigg[Q^{(l),\gamma}_{\beta\alpha}(\tau_2,\tau_1)\bigg],
\end{equation}
where
\begin{align}
\label{Q12}
Q^{(1),\gamma}_{\beta\alpha}(\tau_2,\tau_1)&=\langle \hat{M}_I^{\gamma}(\tau_1+\tau_2)\hat{M}_I^{\beta}(\tau_1)\hat{M}_I^{\alpha}(0)\rangle, \nonumber\\
Q^{(2),\gamma}_{\beta\alpha}(\tau_2,\tau_1)&=-\langle \hat{M}_I^{\beta}(\tau_1)\hat{M}_I^{\gamma}(\tau_1+\tau_2)\hat{M}_I^{\alpha}(0)\rangle,
\end{align}
and
\begin{align}
\chi^{(3),\gamma}_{\beta\alpha\alpha}(\tau_2,\tau_1,0)&=\frac{1}{N}\sum_{l=1}^{4}\Im\bigg[R^{(l),\gamma}_{\beta\alpha\alpha}(\tau_2,\tau_1,0)\bigg],\\
\chi^{(3),\gamma}_{\beta\beta\alpha}(\tau_2,0,\tau_1)&=\frac{1}{N}\sum_{l=1}^{4}\Im\bigg[R^{(l),\gamma}_{\beta\beta\alpha}(\tau_2,0,\tau_1)\bigg],
\end{align}
in which
\begin{align}
\label{R_l_cor_function}
R^{(1),\gamma}_{\beta\eta\alpha}(t_3,t_2,t_1)&=\langle \hat{M}_I^{\eta}(t_1)\hat{M}_I^{\beta}(t_1+t_2)\hat{M}_I^{\gamma}(t_1+t_2+t_3)\hat{M}_I^{\alpha}(0)\rangle,\nonumber\\
R^{(2),\gamma}_{\beta\eta\alpha}(t_3,t_2,t_1)&=\langle \hat{M}_I^{\alpha}(0)\hat{M}_I^{\beta}(t_1+t_2)\hat{M}_I^{\gamma}(t_1+t_2+t_3)\hat{M}_I^{\eta}(t_1)\rangle,\nonumber\\
R^{(3),\gamma}_{\beta\eta\alpha}(t_3,t_2,t_1)&=\langle \hat{M}_I^{\alpha}(0)\hat{M}_I^{\eta}(t_1)\hat{M}_I^{\gamma}(t_1+t_2+t_3)\hat{M}_I^{\beta}(t_1+t_2)\rangle,\nonumber\\
R^{(4),\gamma}_{\beta\eta\alpha}(t_3,t_2,t_1)&=\langle \hat{M}_I^{\gamma}(t_1+t_2+t_3)\hat{M}_I^{\beta}(t_1+t_2)\hat{M}_I^{\eta}(t_1)\hat{M}_I^{\alpha}(0)\rangle,
\end{align}
where $\eta=\alpha$ or $\beta$ and the subscript $I$ means the magnetization is calculated in the interaction picture.  As we expect from the sequence of the  magnetic impulses and  the  recorded signal  after them, in the interaction picture the magnetizations $ \hat{M}_I^{(\alpha)} $, $ \hat{M}_I^{(\beta)} $ and $  \hat{M}_I^{(\gamma)} $ always appear at times $ (0),\ (t_1+t_2)$, and $(t_1+t_2+t_3 )$, respectively.

\section{The model} \label{Model}
The extended Kitaev model describes spin-1/2 degrees of freedom on the honeycomb lattice that is composed of the pure Kitaev terms along with three spin interactions.  This model is an effective theory
-- which is obtained by a perturbative expansion \cite{Kitaev_2006}-- to explain the effect of a weak magnetic field on Kitaev model,
\begin{equation}
\label{model_Ham}
H_{EK}=-\sum_{\langle ij\rangle_{\alpha}}J_{\alpha}\hat{\sigma}^{\alpha}_i\hat{\sigma}^{\alpha}_j-K\sum_{\substack{\langle ik\rangle_{\alpha},\langle kj\rangle_{\beta}\\\gamma\perp\alpha,\beta}}\hat{\sigma}^{\alpha}_i\hat{\sigma}^{\gamma}_k\hat{\sigma}^{\beta}_j.
\end{equation}
In the last term the two bonds 
$ \langle ik\rangle_{\alpha} $ and 
$\langle kj\rangle_{\beta}  $ share the common site k. 
Following  Refs.~[\onlinecite{Pedrocchi_Loss}], we 
will consider the same labelling of spins and boundary conditions for the lattice, that is a honeycomb lattice with  periodic boundary conditions in the direction of two base vectors 
$ L_1\boldsymbol{e}_1 $ and $ L_2\boldsymbol{e}_2+M\boldsymbol{e}_1 $, see Fig.~\ref{lattice}-(a). In this geometry, the number of unit cells is $N=L_1L_2$.

For each plaquette $p$, Fig.~\ref{lattice}-(a), the product of spins sitting on the corners is a constant of motion: 
$ \hat{W}_p=\hat{\sigma}^x_1\hat{\sigma}^y_2\hat{\sigma}^z_3\hat{\sigma}^x_4\hat{\sigma}^y_5\hat{\sigma}^z_6$, which commutes with Hamiltonian and with other plaquette operators
$[\hat{W_p}, \hat{W_{p'}}]=0$. Using the Kitaev parton construction,  each spin is constructed   with  four Majorana fermions, the three static $\hat{b}_i^x,\hat{b}_i^y,$ and $\hat{b}_i^z $ and the dynamic  $ \hat{c}_i$ as follows\cite{Kitaev_2006}:
\begin{equation}
\hat{\sigma}_i^{\alpha}=i\hat{b}_i^{\alpha}\hat{c}_i,\quad\alpha=x,y,z.
\end{equation}

Using this representation we can rewrite the Hamiltonian ($H_{EK}$) in terms of Majorana fermions,
\begin{eqnarray}
\label{first_u_second_uu}
H_{EK}(\{u_{ij}\})=\frac{i}{2}\sum_{\langle ij\rangle_{\alpha}}J_{\alpha}\hat{u}_{\langle ij\rangle_{\alpha}}\hat{c}_i\hat{c}_j&-&\frac{i}{2}K\sum_{\substack{\langle ik\rangle_{\alpha},\langle jk\rangle_{\beta}\\\gamma\perp\alpha,\beta}}\epsilon_{\alpha\gamma\beta}\hat{u}_{\langle ik\rangle_{\alpha}}\hat{u}_{\langle jk\rangle_{\beta}}\hat{c}_i\hat{c}_j,\nonumber\\
\hat{u}_{\langle ij\rangle_{\alpha}}&=&i\hat{b}^{\alpha}_i\hat{b}^{\alpha}_j,
\label{H_MF}
\end{eqnarray}
where $  \epsilon_{\alpha\gamma\beta} $ is the Levi-Civita symbol. Since 
$[\hat{H}_{EK},\hat{u}_{\langle ij\rangle_{\alpha}}]= 0$ and 
$[\hat{u}_{\langle ij\rangle_{\alpha}},\hat{u}_{\langle ij\rangle_{\beta}}]=0$, for a given set of the bond variables $\{{u}_{\langle ij\rangle_{\alpha}}=\pm1\}$, the Hamiltonian is reduced to a hopping problem of Majorana fermions, which can be solved exactly. $ u_{ij} $ is an emergent $\mathbb{Z}_2$ gauge field, which makes the Hilbert space to be factorized into gauge $ |\mathcal{G}\rangle $ and matter 
$ |\mathcal{M}\rangle $ sectors. 
The physics of  the spin Hamiltonian $H_{EK} $  is determined by the flux configurations 
$ \{W_p=\prod_{\langle ij\rangle\in\partial p}u_{ij}\} $, where  different gauge (bond) configurations $ \{u_{ij}\} $ 
could give the same flux configuration. At a fixed flux sector the Hamiltonian takes the following compact form:\cite{Knolle_Chalker_Moessner2015}
\begin{equation}
\label{H_u_c}
H_{EK}=\frac{i}{2}\begin{pmatrix} 
\hat{c}_A^T & \hat{c}_B^T 
\end{pmatrix}
\begin{pmatrix} 
F & M \\
-M^T& -G
\end{pmatrix}
\begin{pmatrix} 
\hat{c}_A\\
\hat{c}_B
\end{pmatrix},
\end{equation}
where $ \hat{c}_{A (B)}  $ is the column  vector of all matter Majorana fermions on A (B) sublattice.  $ M $ is the first neighbor hopping matrix, while $ F $ and $ G $ are the second neighbor hopping matrices. In order to diagonalize  $H_{EK}(\{u_{ij}\})$ in each flux sector, we  introduce complex gauge and matter fermions that act on matter and gauge sectors, respectively \cite{Baskaran2007}:
\begin{align}
\label{complex_fermion}
&\hat{f}_{\mu_z}=\frac{1}{2}(\hat{c}_{\mu A}+i\hat{c}_{\mu B}),\quad
\hat{\chi}_{\mu_z}=\frac{1}{2}(\hat{b}^z_{\mu A}-i\hat{b}^z_{\mu B}),\nonumber\\
&\hat{\chi}_{\mu_y}=\frac{1}{2}(\hat{b}^y_{\mu A}-i\hat{b}^y_{\mu' B}),\quad
\hat{\chi}_{\mu_x}=\frac{1}{2}(\hat{b}^x_{\mu A}-i\hat{b}^x_{\mu'' B}).
\end{align}
According to our notation, $ \mu $ labels the unit cells and $\mu_{a}, a=x, y, z$  indicates  the $x/y/z$-bond  in that unit cell,
which are shown in Fig.~\ref{lattice}-(a). 
The gauge configuration $ \{u_{ij}\} $, is determined by the occupation number of gauge fermions using the relation: $ \hat{u}_{\langle ij\rangle_{\alpha}}=1-2\hat{\chi}^{\dagger}_{\langle ij\rangle_{\alpha}}\hat{\chi}^{}_{\langle ij\rangle_{\alpha}} $.
Hence, the Hamiltonian in each flux sector in terms of  complex fermions ($\hat{f}$) takes this form:\cite{Knolle_Chalker_Moessner2015}
\begin{equation}
\label{H_complex}
H_{EK}=\frac{1}{2}\begin{pmatrix} 
\hat{f}^{\dagger}&\hat{f}\end{pmatrix}
\begin{pmatrix} 
h&\Delta\\
\Delta^{\dagger}& -h^{*}
\end{pmatrix}
\begin{pmatrix} 
\hat{f}\\\hat{f}^{\dagger}\end{pmatrix},
\end{equation}
where
\begin{align}
h&=(M^T+M)+i(F-G),\quad
h^{\dagger}=h
\nonumber\\
\Delta&=(M^T-M)+i(F+G),\quad \Delta^T=-\Delta.
\end{align}

Using the Bogoliubov  transformation, $ U $, as has been described in  Refs.~[\onlinecite{Knolle_Chalker_Moessner2015},\onlinecite{Blaizot}], the final Hamiltonian is diagonalized as follows
\begin{equation}
\label{eigenstates}
H_{EK}=\sum_{n}\varepsilon_n\hat{a}^{\dagger}_n\hat{a}_n-\frac{1}{2}\sum_n\varepsilon_n,
\end{equation}
where
\begin{equation}
\label{U_transp}
\begin{pmatrix}
\hat{a}\\
\hat{a}^{\dagger}
\end{pmatrix}=
U\begin{pmatrix}
\hat{f}\\
\hat{f}^{\dagger}
\end{pmatrix}.
\end{equation}
Here,   $\varepsilon_n\geq0$ is the matter excitation
energy and  $ \hat{a}^{\dagger}_n $($ \hat{a}_n $) is the canonical  fermionic creation (annihilation) operator. The last term in Eq.(\ref{eigenstates}), 
\begin{equation}
E=-\frac{1}{2}\sum_n\varepsilon_n, 
\label{gse}
\end{equation}
is the ground state energy (i.e., without matter excitations) in each flux sector. 

\subsection{Projection operator and the physical states}

Representation of a spin with four Majorana operators has doubled the dimension of Hilbert space on each site of lattice\cite{moessner_moore_2021}.
 Therefore, not all states in the extended Hilbert space $(\mathcal{H}_{EK})$ belong to the original physical spin Hilbert space. The states in the extended Hilbert space can be classified into physical and unphysical states by 
 introducing  the projection operator $\hat{\mathcal{P}}$ \cite{Kitaev_2006},
\begin{align}
&\hat{\mathcal{P}}=\prod_{i=1}^{2N}\bigg(\frac{1+\hat{D}_i}{2}\bigg),\quad\text{with:}\quad \hat{D}_i=\hat{b}_i^{x}\hat{b}_i^{y}\hat{b}_i^{z}\hat{c}_i,\nonumber\\
&|\Psi_{\text{phys}}\rangle=\hat{\mathcal{P}}|\Psi_u\rangle.
\end{align}

Within a straightforward calculation we find that the projection operator can be written in the following form\cite{Yao_Kivelson_2009,Yao_Qi_2010}
\begin{align}
&\hat{\mathcal{P}}= \hat{\mathcal{S}} \hat{\mathcal{P}}_0,\nonumber\\
& \hat{\mathcal{P}}_0=\frac{1+\hat{D}}{2},
\end{align}
where $\hat{D}=\prod_{i=1}^{2N}\hat{D}_i$ and $\hat{\mathcal{S}}$ sums symmetrically over all gauge-equivalent $\{u_{ij}\}$ configurations. For physical (unphysical) states we have $D=+1(-1)$.  
It has been shown that the operator $\hat{D}$ depends on the following values and parities \cite{Pedrocchi_Loss}:
\begin{equation}
\label{parity_proj}
\hat{D}=(-1)^{\theta}\det(Q_u)\hat{\pi}_{\chi}\hat{\pi}_a
\end{equation}
where $\theta=L_1+L_2+M(L_1-M)$ and $Q_u$ is obtained  by diagonalizing the Hamiltonian, see Appendix~\ref{projection_oper}.  Moreover, $\hat{\pi}_{\chi}$ and $\hat{\pi}_{a}$ are the number parity of bond and matter fermions.

\subsection{Physical ground state}
\label{phy_unphy_states}

The Wilson loop operator $ \hat{W}_{\Gamma} $ associated with any closed loop $ \Gamma $ on the lattice is a constant of motion for the Hamiltonian (\ref{model_Ham})\cite{Kitaev_2006,Halasz_Chalker_2014}: 
\begin{equation}
\hat{W}_{\Gamma}=\hat{\sigma}_{i}^{\alpha_{\langle ij\rangle}}\hat{\sigma}_j^{\alpha_{\langle ij\rangle}}
\hat{\sigma}_{j}^{\alpha_{\langle jk\rangle}}
\hat{\sigma}_{k}^{\alpha_{\langle jk\rangle}}\dots
\hat{\sigma}_{l}^{\alpha_{\langle li\rangle}}
\hat{\sigma}_{i}^{\alpha_{\langle li\rangle}},
\end{equation}
where  $ \{i,j,k,\dots,l\} $ refer to the sites on the loop and $ \alpha_{\langle ij\rangle}=x,y,z$ shows the type of  connecting link $\langle ij\rangle_{\alpha}  $, i.e., the Wilson loop is the product of $ x/y/z- $ Ising interactions on  $ x/y/z- $ links of the loop.
The flux (plaquette) operator $ \hat{W}_p $ is an elementary closed-loop operator such that any contractible closed-loop operator $\hat{W}_{\Gamma}$ on a torus can be constructed by multiplying a sequence of $ \hat{W}_p $. On a torus (2D lattice with periodic boundary conditions),  there are two non-contractible (topological) closed-loop operators that can not be construed by the product of plaquette operators.  For example, in a system with the boundary condition $M=0$, these loop operators are $ \hat{W}_{L_1} $ and $ \hat{W}_{L_2} $ as shown with the red links in Fig.\ref{lattice}-(a,b). The eigenvalues of these operators are $ l_1=\pm1 $ and $ l_2=\pm1 $. So, for any flux configuration $ \{W_p\} $, there are four topologically inequivalent states, namely  $ |\{W_p\},l_1,l_2\rangle$.

\begin{figure}[t]
	\center
	\includegraphics[scale=0.7]{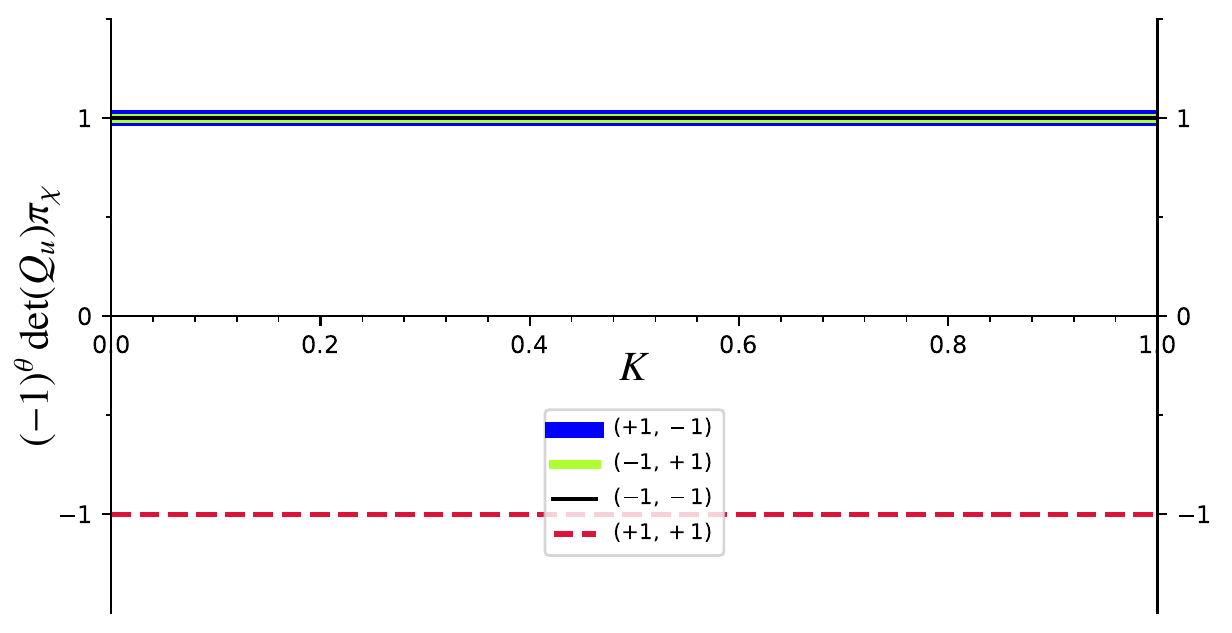}
	\caption{The value of $(-1)^{\theta}\det(Q_u)\hat{\pi}_{\chi}$ versus $K$ for the four topological
	distinct states introduced in Eq.(\ref{topological_states}) labelled by ($l_1, l_2$). The exchange couplings are in the non-Abelian phase
	with $ J_x=J_y=J_z=1$ for a periodic system of $ L_1=L_2=56, M=0 $.}
	\label{sign_det_fig}
\end{figure}

According to the Lieb's theorem\cite{Lieb73}, we look for the physical ground state 
$(D=+1)$, in the 0-flux sector $ \{W_p=+1, \forall p\} $. For simplicity, we consider  a system with $L_1=L_2$ being an even number and $M=0$.  Four topologically inequivalent states for the 0-flux sector can be constructed with the following gauge configurations:
\begin{align}
\label{topological_states}
&|\{W_p=+1\},+1,+1\rangle=\hat{\mathcal{S}}|\mathcal{G}\rangle|\mathcal{M'^{++}}\rangle,\nonumber\\
&|\{W_p=+1\},+1,-1\rangle=\hat{\mathcal{S}}g_{+-}(\hat{\chi}^{\dagger})|\mathcal{G}\rangle|\mathcal{M'}^{+-}\rangle,
\nonumber\\
&|\{W_p=+1\},-1,+1\rangle=\hat{\mathcal{S}}g_{-+}(\hat{\chi}^{\dagger})|\mathcal{G}\rangle|\mathcal{M'}^{-+}\rangle,
\nonumber\\
&|\{W_p=+1\},-1,-1\rangle=\hat{\mathcal{S}}g_{--}(\hat{\chi}^{\dagger})|\mathcal{G}\rangle|\mathcal{M'}^{--}\rangle,
\end{align}
where $ |\mathcal{G}\rangle $ is the vacuum for the complex gauge fermions defined by the standard gauge configuration $ \{u_{ij}=+1\} $  and $ |\mathcal{M^{++}}\rangle $ is  the vacuum for matter fermions in this gauge configuration. The operator $g_{l_1l_2}(\hat{\chi}^{\dagger}) $ is the product of creation operators for complex gauge fermions which construct the gauge configuration with a specific topological label $(l_1,l_2)$ and $|\mathcal{M}^{l_1l_2}\rangle$ is the vacuum for matter fermions in the aforementioned gauge configuration $  g_{l_l l_2}(\hat{\chi}^{\dagger})|\mathcal{G}\rangle $. The prime on the matter state $ |\mathcal{M'}^{l_1l_2}\rangle $ indicates that this state  may have  matter excitations, which is depends on the factor $(-1)^{\theta}\det(Q_u)\hat{\pi}_{\chi}$  defined in Eq.(\ref{parity_proj}). 
We have plotted the value of $(-1)^{\theta}\det(Q_u)\hat{\pi}_{\chi}$ versus $K$ in 
Fig.~\ref{sign_det_fig} for
the  gauge configurations defined in Eq.(\ref{topological_states}) at the isotropic point $J_x=J_y=J_z =1$. According to Eq.(\ref{parity_proj}) and Fig.~\ref{sign_det_fig}, to reach a physical state with $D=+1$ 
the state with topological label $ (+1,+1) $ must have an odd parity for matter excitations, while
the other topological ground states have zero matter excitation. 
Based on our numerical evidences, we expect that for all even and odd values of the geometric parameters $L_1 $, $ L_2$, and  $M$, the odd parity constraint for physical states with label $ (+1,+1) $ in the 0-flux sector  holds  in the entire area of non-abelian phase of the extended Kitaev model, similar to the pure Kitaev model\cite{Zschocke_Vojta}. Accordingly, the states in Eq.(\ref{topological_states}) with  minimum energy must have the following matter configurations,
\begin{align}\label{tp-groundstates}
&|\mathcal{M'^{++}}\rangle=\hat{a}^{\dagger}_1|\mathcal{M^{++}}\rangle\nonumber\\  
&|\mathcal{M'^{+-}}\rangle=|\mathcal{M^{+-}}\rangle, \nonumber\\
&|\mathcal{M'^{-+}}\rangle=|\mathcal{M^{-+}}\rangle, \nonumber\\
&|\mathcal{M'^{--}}\rangle=|\mathcal{M^{--}}\rangle.
\end{align}
It means the energy of $|\{W_p=+1\},+1,+1\rangle$ is $ E^{(0)}+\varepsilon_1^{(0)} $, while
the energy of the other three states is
$ E^{(0)} $, which is given by Eq.(\ref{gse}) within the 0-flux sectors and $ \varepsilon_1^{(0)} $ is the first matter excitation in the same flux sector. Given $\varepsilon_1^{(0)} > 0 $  for any finite and non-zero value of $ K $, the energy of the 0-flux state  with the  label $ (+1,+1) $ is higher by the fermionic gap than the other three states in the groundstate manifold.  
Therefore, the topological groundstate in the non-Abelian phase of the extended model is three-fold degenerate 
as defined in Eq.(\ref{tp-groundstates}) in agreement with the results presented in
Ref.~[\onlinecite{Kells_2009}] using the Jordan-Wigner type transformation.
It has to be mentioned that in  Ref.~[\onlinecite{Kells_2009}] the transformation is in the original Hilbert space of the model and there is no unphysical degrees of freedom. 
Moreover,  the non-Abelian phase supports three types of quasiparticles, namely: vacuum, Ising anyons, and fermions. 
Hence,  
in the framework of topological quantum field theory the groundstate on a torus has three-fold degeneracy\cite{Nayak_2008}. In the Abelian phase of the model, we observed that for even value of $L_1$ and  $L_2$ with $M=0$, the factor $ (-1)^{\theta}\det(Q_u)\hat{\pi}_{\chi}$
is always equal to $+1$ for all topologically inequivalent states, i.e, in this case, $ |\mathcal{M'^{++}}\rangle=|\mathcal{M^{++}}\rangle $ and the  groundstate subspace is composed of four degenerate states.

%%%%%%%%%%%%%%%%%%%%%%%%%%%%%%%%
%%%%%%%%%%%%%%%%%%%%%%%%%%%%%%%%

In order to calculate the nonlinear response of the system in the Abelian and non-Abelian phases, we can choose any state from the ground state manifold, because for 2DCS as a local probe, topologically inequivalent ground states are indistinguishable.

%%%%%%%%%%%%%%%%%%%%%%%%%%%%%%%%%%%%%%%%%%%%%%%%%%%%%%%%%%%

%%%%%%%%%%%%%%%%%%%%%%%%%%%%%%%%%%%%%%%%%%%%%%%%%%%%%%%%%%%%%%

\section{The nonlinear response}\label{results}
\begin{figure*}[!htb]
	\centering
	\includegraphics[scale=1.3]{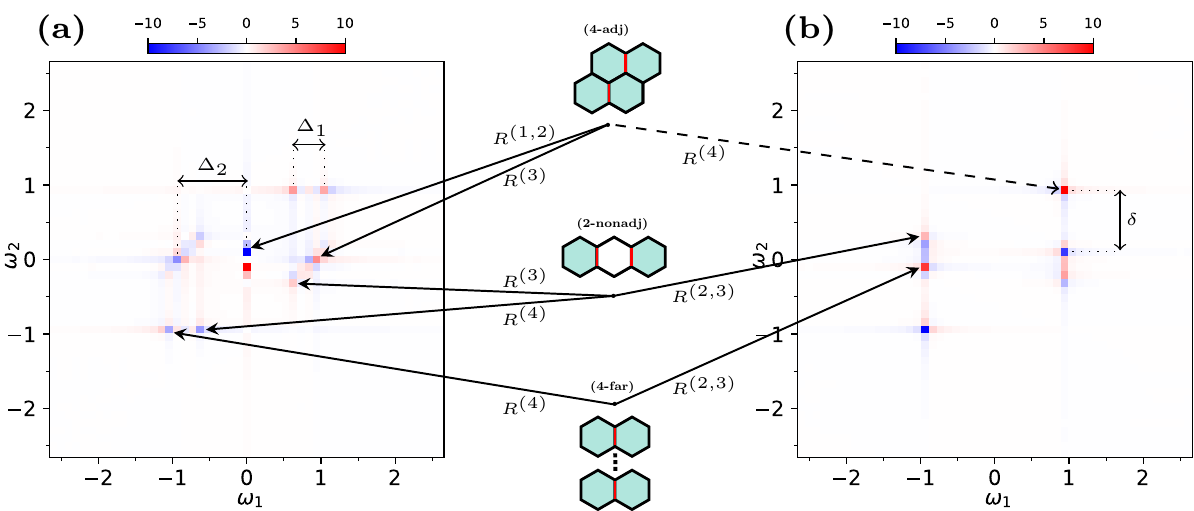}
	\caption{Two-dimensional nonlinear response in frequency domain at the isotropic point ($J_x=J_y=J_z=1$) in the non-Abelian phase with $K=0.2$. (a) $\chi^{(3),z}_{zzz}(\omega_2,\omega_1,0)$. (b) $\chi^{(3),z}_{zzz}(\omega_2,0,\omega_1)$. The dashed arrow indicates that the main contribution to the signal magnitude comes from the corresponding flux states. The solid arrows indicate that the signal comes only from the corresponding states.  The black dots in  the (4-far) class in the middle panel means that the flux separation is equal or more than two plaquettes. The maximum signal peak is normalized to 10.}
	\label{2FT_phy}
\end{figure*}

The two-dimensional nonlinear susceptibilities introduced in Eq.(\ref{non_lin_formula}) is 
calculated for the extended Kitaev model at polarization $(\alpha,\beta,\gamma)=(z,z,z)$.
For the mentioned polarization, the second-order susceptibility is exactly zero, 
because by insertion of the resolution identity  in the  correlation functions $ Q^{(l),\gamma}_{\beta\alpha} $,  the overlap of flux sectors of the intermediate states vanishes \cite{choi1}.
We have used the few matter fermion approach, specifically the single matter fermion approximation \cite{Knolle_Chalker_Moessner2014}; which shows that the dynamical structure factor for small $ K\text{'s} $ within this approximation gives almost the same results as the exact Paffian approach. To elaborate on this, it suffices to  insert the resolution of identity into $R^{(1),\gamma}_{\beta\eta\alpha}$ correlation functions. For example, consider $R^{(1),z}_{zzz}(\tau_2,\tau_1,0)=R^{(2),z}_{zzz}(\tau_2,\tau_1,0)\doteq R^{(1,2),z}_{zzz}(\tau_2,\tau_1,0) $:
\begin{align}
\label{R_12_general}
&R^{(1,2),z}_{zzz}(\tau_2,\tau_1,0)=\langle \hat{M}_I^{z}(0)\mathds{1}\hat{M}_I^{z}(\tau_1)\mathds{1}\hat{M}_I^{z}(\tau_1+\tau_2)\mathds{1}\hat{M}_I^{z}(0)\rangle\nonumber\\
&=\sum_{\mu\nu\lambda\rho}\sum_{PQR}\langle G|\hat{Z}_{\mu}|P\rangle \langle P|\hat{Z}_{\nu}|Q\rangle \langle Q|\hat{Z}_{\lambda}|R\rangle \langle R|\hat{Z}_{\rho}|G\rangle\nonumber\\
&\quad\quad\quad\quad\times e^{i[E_P\tau_1+E_Q\tau_2-E_R(\tau_1+\tau_2)]},
\end{align}
where $ \hat{Z}_{\mu}=\hat{\sigma}^{z}_{\mu A}+\hat{\sigma}^{z}_{\mu B} $, is the sum of the Pauli z-matrix of the two spins 
on the $\mu$-th cell. Moreover, $|G\rangle$ is the ground state, $|P\rangle$, $|Q\rangle$ and $|R\rangle$ represent an eigenstate of the 
Hamiltonian.
The  matrix elements appeared in Eq.(\ref{R_12_general}) are the same for other $R^{(1),z}_{zzz} $ functions, which differ only in the phase factor. The  eigenstates with different flux sectors are orthogonal to each other, 
hence 
the matrix elements in Eq.(\ref{R_12_general}) are non-zero  
only for the four $\hat{Z}_{\mu}$ operators with the following sequence of indices: 
$\hat{Z}_{m}\hat{Z}_{n}\hat{Z}_{n}\hat{Z}_{m}$, 
$\hat{Z}_{m}\hat{Z}_{n}\hat{Z}_{m}\hat{Z}_{n}$, and, 
$\hat{Z}_{m}\hat{Z}_{m}\hat{Z}_{n}\hat{Z}_{n}$. The last case, in the single matter approximation produces a term, which grows linearly with the system size.  However, according to Eq.(\ref{non_lin_formula}), the
nonlinear susceptibilities are independent of the system size,
hence, we exclude this contribution\cite{choi1}.   The two former cases  result in the following matrix elements,
\begin{align}
\label{phy_contributions}
\sum_{\mu\nu}\langle G|\hat{Z}_{\mu}|P_{\mu}\rangle \langle P_{\mu}|\hat{Z}_{\nu}|Q_{\mu\nu}\rangle \langle Q_{\mu\nu}|\hat{Z}_{\nu}|R_{\mu}\rangle \langle R_{\mu}|\hat{Z}_{\mu}|G\rangle,\nonumber\\
\sum_{\mu\nu}\langle G|\hat{Z}_{\nu}|P_{\nu}\rangle \langle P_{\nu}|\hat{Z}_{\mu}|Q_{\mu\nu}\rangle \langle Q_{\mu\nu}|\hat{Z}_{\nu}|R_{\mu}\rangle \langle R_{\mu}|\hat{Z}_{\mu}|G\rangle.
\end{align}
According to the single matter approximation we choose  the states in Eq.(\ref{phy_contributions}) as follows:
\begin{equation}
\label{phy_states_non_abelian}
\begin{cases}
|G\rangle=\hat{\mathcal{S}}g_{+-}(\hat{\chi}^{\dagger})|\mathcal{G}\rangle|\mathcal{M}^{+-}\rangle\\
|{R_{\mu}}\rangle=\hat{\mathcal{S}}\hat{\chi}_{\mu_z}^{\dagger}g_{+-}(\hat{\chi}^{\dagger})
\overline{\hat{a}}_r^{\dagger}
|{\mathcal{G}}\rangle|\mathcal{M}_{\mu}^{+-}\rangle\hspace{8mm} (\mbox{2-flux state})\\
|{Q_{\mu\nu}}\rangle=\hat{\mathcal{S}}\hat{\chi}_{\mu_z}^{\dagger}
\hat{\chi}_{\nu_z}^{\dagger}g_{+-}(\hat{\chi}^{\dagger})
|{\mathcal{G}}\rangle|\mathcal{M}_{\mu\nu}^{+-}\rangle \hspace{5mm}\  (\mbox{2 or 4-flux state}),
\end{cases}
\end{equation}
where $\overline{\hat{a}}_r^{\dagger}$ denotes a fermion creation operator in the 2-flux state $ |R_{\mu}\rangle $  and $  |\mathcal{M_{\mu}^{+-}}\rangle$, $  |\mathcal{M_{\mu\nu}^{+-}}\rangle$ are the vacuum for matter fermions. 
In our calculations, we consider the state with topological label $ (+1,-1) $ for the ground state of nonlinear response in the Abelian and non-Abelian phases.

%%%%%%%%%%%%%%%%%%%%%%%%%%%%%%%%%%%%%
\begin{figure*}[t]
	\centering
	\includegraphics[scale=1.45]{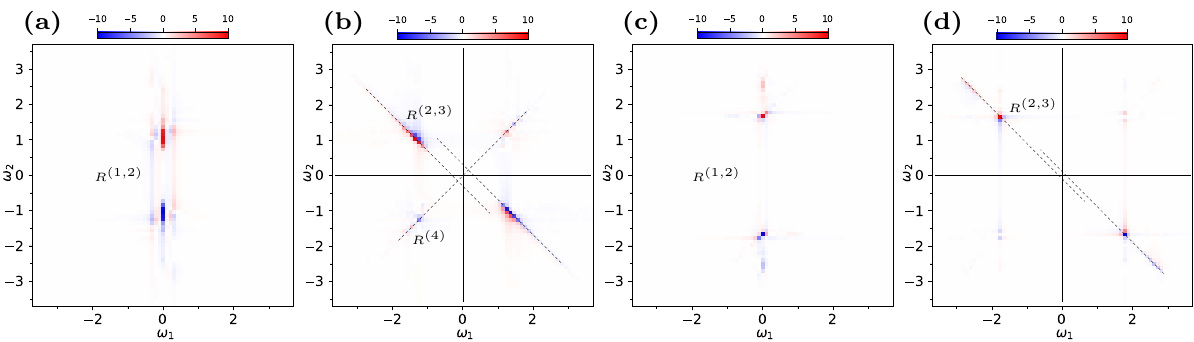}
	\caption{Two-dimensional nonlinear susceptibilities in frequency domain at an anisotropic point in the Abelian phase with $ K=0.2 $.  (a) $ \chi^{(3),z}_{zzz}(\omega_2,\omega_1,0) $ and (b)  $\chi^{(3),z}_{zzz}(\omega_2,0,\omega_1)$ with $J_x =1, J_y = J_z = 0.3$. (c) $ \chi^{(3),z}_{zzz}(\omega_2,\omega_1,0) $ and (d)  $\chi^{(3),z}_{zzz}(\omega_2,0,\omega_1)$ with $J_x =1, J_y = J_z = 0.05$.  In (b) and (d) the shifted diagonal signals intercept $ \omega_1 $ and $ \omega_2 $ axes at $ |E_4-E_0|=0.276$ and $ 0.125 $, respectively. 
		The magnitude of the maximum peak of susceptibility in all plots is normalized to 10.}
	\label{2FT_abelain}
\end{figure*}
%%%%%%%%%%%%%%%%%%%%%%%%%%%%%%%%%%%%

\subsection{Results}\label{nonlinaer_result}
The Fourier transform of two-dimensional nonlinear magnetic susceptibilities, $\chi^{(3),z}_{zzz}(\omega_2,\omega_1,0)$ and $\chi^{(3),z}_{zzz}(\omega_2,0,\omega_1)$, are shown in Fig.~\ref{2FT_phy} for a system with ($33\times33$) unit cells, $J_x=J_y=J_z=J=1$ and $K=0.2$.
We use the fast Fourier transformation to obtain the Fourier spectrum from time domain\cite{comment3}. 
In the non-Abelian phase of the model, the energy of a flux state depends on the distance between its vortices\cite{lahtinen2008,Lahtinen_Interacting_2011},  accordingly the  states $ |Q_{1\mu} \rangle$ in  the nonlinear response  can be classified into three classes:  4-adjacent, 2-non-adjacent, 4-far-separated, as depicted in the middle panel of Fig.~\ref{2FT_phy}.
By considering the contribution of states in each of these classes individually, we can determine the origin of sharp peaks in the observed response. The origin of each peak in the  response in terms of the mentioned classes are indicated by the corresponding  arrows in the middle panel of Fig.~\ref{2FT_phy}. 
A black arrow pointing to a spot in response stipulates the origin of that signal comes only from the mentioned 
excitations, while the dashed  arrow indicates an excitation which has the main contributions to the signal.
The observation of peaks corresponding to flux sectors with 4-adjacent, 2-non-adjacent, and 4-far-separated fluxes is {\it a new signature of flux excitations that does not appear in the linear responses}. Due to the fact that the extended model has a gapped spectrum, a system with $ (33\times 33) $ unit cells is large enough to reach the thermodynamic limit for $K=0.2$. The finite size effects is discussed in the appendix (See Fig.(\ref{finite_size})), where we present the difference between physical responses in Fourier space for different system sizes. As the system size increases to $L=33$, the aforementioned difference becomes almost zero, which convinces us to reach
enough large system sizes.
The three energy scales $ \Delta_1 $, $ \Delta_2 $, and $ \delta $ can be extracted from the nonlinear response, as depicted in Fig.~\ref{2FT_phy},
\begin{align}
\label{delta_enegrey}
&\Delta_1=(E^{(4\text{far})}-E^{(0)})-(E^{(2\text{nonadj})}-E^{(0)})=E^{(4\text{far})}-E^{(2\text{nonadj})},\nonumber\\
&\Delta_2=E^{(4\text{adj})}-E^{(0)},\nonumber\\
&\delta=(\varepsilon_1^{(2)}+E^{(2)}-E^{(0)})-(E^{(4\text{far})}-E^{(2)}-\varepsilon_1^{(2)})\approx  2\varepsilon_{1}^{(2)},\nonumber\\
\end{align}
where $E^{(4\text{adj})}$, $ E^{(2\text{nonadj})}$, and $E^{(4\text{far})}$ are the ground state energy of the flux states depicted in  the middle of Fig.~\ref{2FT_phy}.  Moreover,  $\varepsilon_{1}^{(2)}$ is  the in-gap energy of two-adjacent flux state.
For the considered parameters in Fig. ~\ref{2FT_phy} we obtain  $( \Delta_1,\Delta_2,\delta)=(0.434,0.894,0.758)$. 
The signature of an in-gap bound state $ \delta $  can be also detected in the linear response as a sharp peak in the dynamical spin structure factor\cite{Knolle_Chalker_Moessner2015},  while $ \Delta_1 $ and $ \Delta_2 $ are {\it new signature of flux states} of the non-Abelian anyons, which appear only in the {\it nonlinear responses}. 

We have also obtained the nonlinear response of the Abelian phase. The Fourier transform  of the third-order susceptibilities for the Abelian phase are presented in Fig.~\ref{2FT_abelain}. 
In all plots of Fig.~\ref{2FT_abelain} we keep $J_x=1$ and $K=0.2$,  the couplings in parts (a) and (b) are $J=J_y=J_z=0.3$ and the system has $N=44\times 44$ unit cells, while in parts (c) and (d) we have $J=J_y=J_z=0.05$ and  $N=28\times 28 $ unit cells.

We observe weak diagonal signals (the first and third quadrants) and strong shifted diagonal signals (the second and fourth quadrants)  in Fig.~\ref{2FT_abelain}-(b) for $J=0.3$ in $\chi^{(3),z}_{zzz}(\omega_2,0,\omega_1)$.
The diagonal signals come from the $R^{(4),z}_{zzz}(\omega_2,0,\omega_1)$ expression,
in which there are two delta functions with peak frequencies:
\begin{equation}
\omega_1=E_0-E_2-\varepsilon_r^{(2)},\quad\quad
\omega_2=E_0-E_2-\varepsilon_p^{(2)},
\end{equation}
where $\varepsilon_{r/p}^{(2)}$ is the matter excitations in the 2-flux state $ |R\rangle/|P\rangle $. Whenever
$  |R\rangle=|P\rangle $, there is a constructive interference for matrix elements,\cite{choi1} which leads to the diagonal signal (non-rephasing signal), $\omega_1=\omega_2$. Moreover, the expression $ R^{(2,3),z}_{zzz}(\omega_2,0,\omega_1) $ is responsible for the shifted diagonal signals, which contains two  peaks at:
\begin{equation}
\label{R_23_frequence}
\omega_1=E_2-E_0+\varepsilon_p^{(2)},\quad\quad
\omega_2=E_4-E_2-\varepsilon_r^{(2)}.
\end{equation}
This leads to the strong streak signal (rephasing signal) $\omega_1+\omega_2=E_4-E_0$  for $ |R\rangle=|P\rangle $ due to constructive interference. This signal intercepts $\omega_2$-axis at $E_4-E_0$. 

%%%%%%%%%%%%%%%%%%%%%%%%%%%%%%%%%%%%%%%%%
 \begin{figure}[t]
	\centering
	\includegraphics[width=0.9\columnwidth]{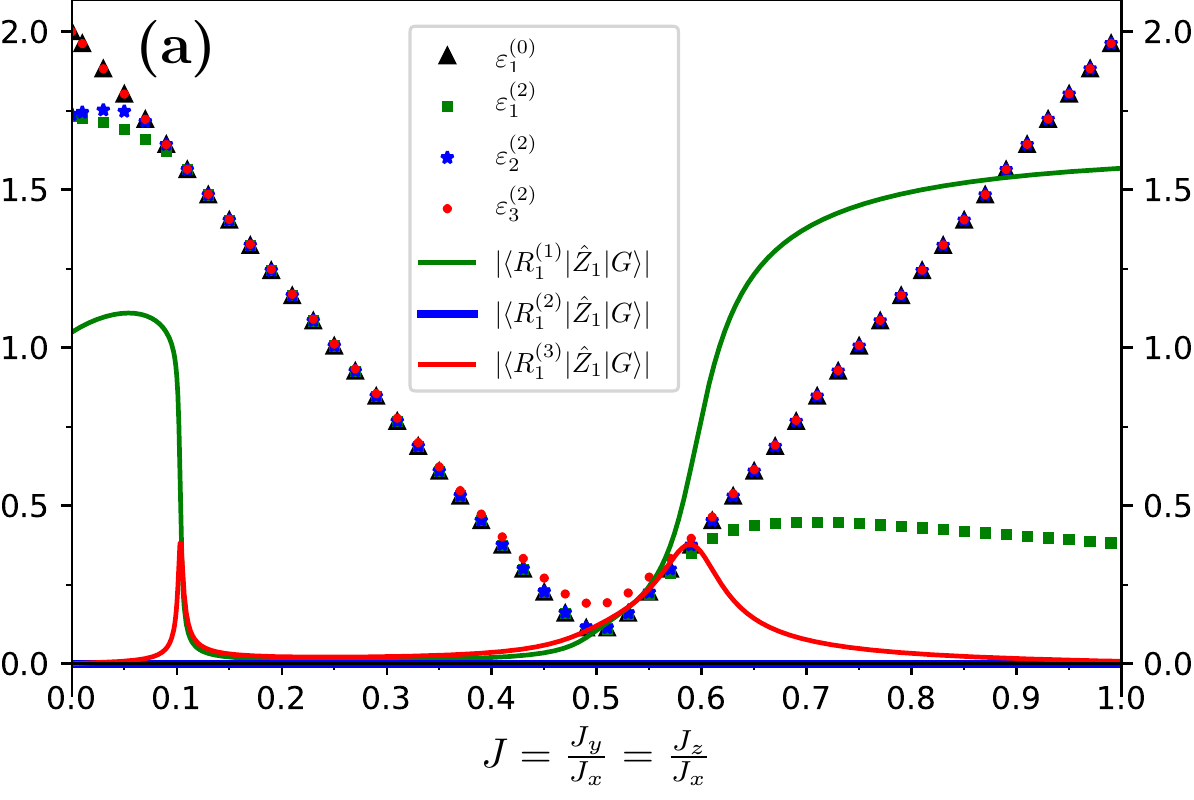}
	\includegraphics[width=0.6\columnwidth]{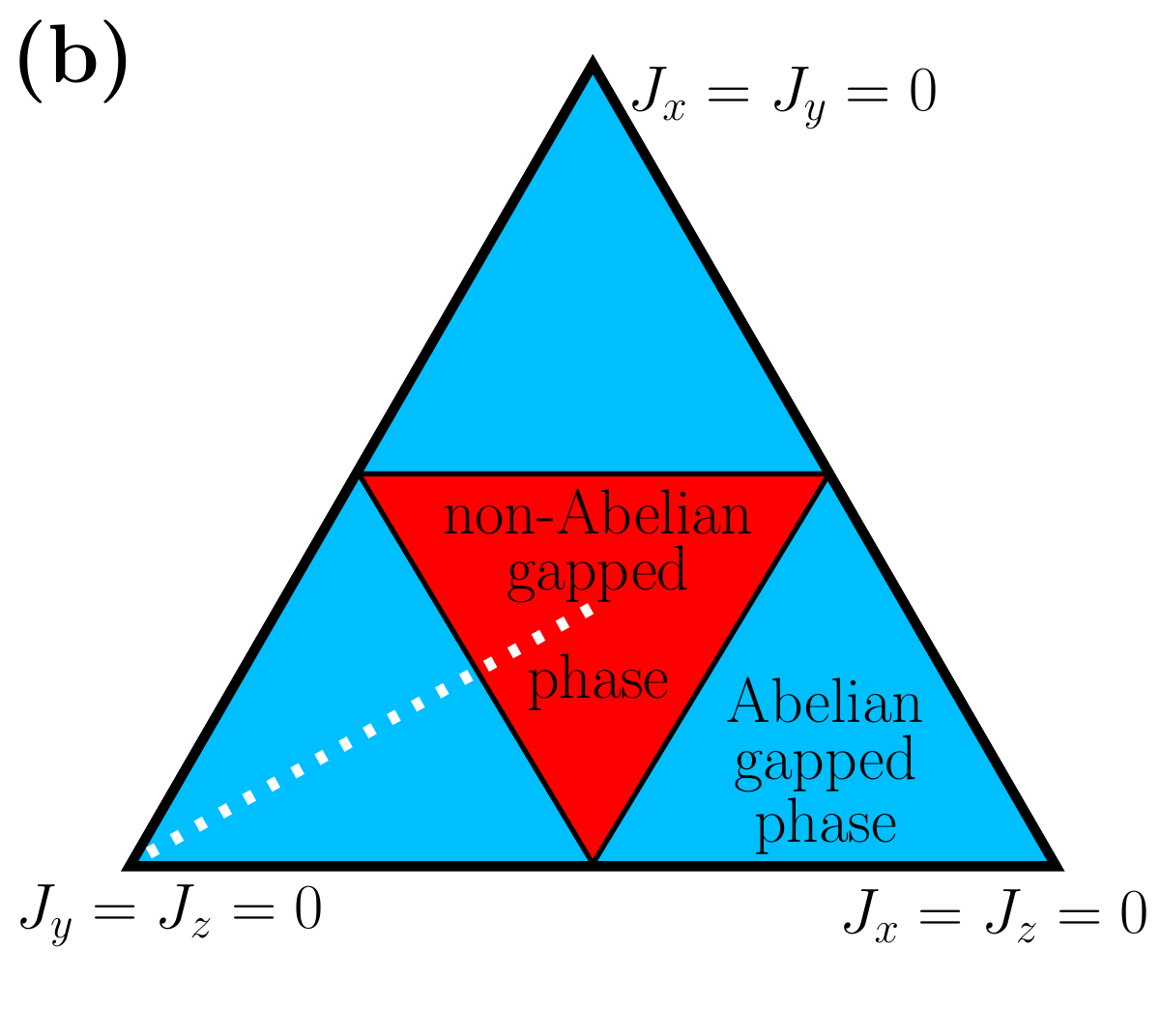}
	\caption{(color online) (a) The evolution path versus $ J $ for the lowest eigenvalue in the 0-flux sector ($ \varepsilon_1^{(0)} $), 2-adjacent flux sector ($\varepsilon_r^{(2)}   $) and the matrix element 
		$|\langle R_{\mu=1}^{(r)}|\hat{Z}_{\mu=1}|G\rangle| $ for the three lowest excitation modes $r=1, 2, 3$ in the two-flux state $ |R_{1}^{(r)}\rangle $. By increasing the size of the system, the  fermionic gap $ \varepsilon_1^{(0)} $ is closed at the phase transition point $ J=0.5 $ in (a). These presented data come from a system with $ L_1=L_2=56,\ M=0 $. (b) The phase diagram of extended Kitaev model in the plane $J_x+J_y+J_z=\mbox{constant}$, where the dotted white line shows the evolution path of part (a). }
	\label{RZG}
\end{figure}
%%%%%%%%%%%%%%%%%%%%%%%%%%%%%%%%%%%%%%%%

However, in Fig.~\ref{2FT_abelain}-(d), at $J=0.05$ the diagonal signal does not show up and the shifted diagonal signal is very weak, where only sharp spots appear in the responses.
 To investigate this difference, we examine 
the absolute value of  $  \langle R_{\mu}|\hat{Z}_{\mu}|G\rangle $ as a relevant matrix element to the nonlinear response versus $J$ within the single matter approximation, 
for the three excited states labelled by $r=1, 2, 3$. The excited states are expressed by 
$|R_{\mu}^{(r)}\rangle=\hat{\mathcal{S}}
\hat{\chi}_{\mu_z}^{\dagger}g_{+-}(\hat{\chi}^{\dagger})
	\overline{\hat{a}}_r^{\dagger}
	|{\mathcal{G}}\rangle|\mathcal{M}^{\mu}_{+-}\rangle$ as  given in Eq.(\ref{phy_states_non_abelian}). 
%Other matrix elements have similar behavior to  $ \langle R_{\mu}^{(r)}|\hat{Z}_{\mu}|G\rangle $.  
A simple expression for this matrix element is given in Appendix(\ref{mat_element}). 
Fig.~\ref{RZG}-(a) shows the first  matter eigenvalue $ \varepsilon^{(0)}_1 $ in the 0-flux state $ |G\rangle $ as well as 
 $ |\langle R_{\mu}^{(r)}|\hat{Z}_{\mu}|G\rangle| $ for $ r=1,2,3 $ corresponding to the lowest  excitation modes $\bar{\hat{a}}^{\dagger}_1  $, $\bar{\hat{a}}^{\dagger}_2$, and $\bar{\hat{a}}^{\dagger}_3  $ in the two-flux state $ |R_{1}^{(r)}\rangle $   and their corresponding eigenvalues $ \varepsilon^{(2)}_1 $, $\varepsilon^{(2)}_2$ , and $ \varepsilon^{(2)}_3 $. 
 The coupling $J$ in Fig.~\ref{RZG}-(a) varies from $ 0 $ to $ 1 $, which is shown by the dotted white path in Fig.\ref{RZG}-(b). 
We anticipate that in the Abelian phase for $ K=0.2, $ the diagonal and shifted diagonal signals to appear in the range $ 0.1< J< 0.5 $,   because low-energy excitation modes have almost equal contributions to the response, which leads to a continuum of spots. However, for $0< J< 0.1$, the excitation mode $ r=1 $ has the dominant matrix element compared with the other excitation modes. Hence, the main  contribution comes from $ r=1 $, and as a result, we only observe sharp spot in the response function. 
According to Fig.~\ref{RZG}-(a), we anticipate that in the non-Abelian phase for $ K=0.2 $ and $ J>0.6 $, the pattern of nonlinear spectrum is formed by sharp spots, similar to what we see in Fig.~\ref{2FT_phy} for the isotropic case.  We have plotted the nonlinear response along the white path of Fig.\ref{RZG}-(b) for several values of $J$ in Fig.\ref{response_evolution} of Appendix \ref{response-J}, which justifies our statement. For details see Appendix \ref{response-J}.

The presence and absence of streak signals in the nonlinear response of the Abelian phase of EKM can also be understood in terms of an effective theory where $J_x \gg J_y, J_z, K$. Deep in the Abelian phase (close to the corners of the triangle in Fig.\ref{RZG}-(b)), the effective Hamiltonian of EKM
is the Kitaev toric code, where the first nonzero term appears in the second order perturbation theory 
(see Appendix \ref{Effective_Hamiltonian}). The toric code has sharp charge (\emph{e})  and flux (\emph{m})  excitations without any dispersion. Therefore, it is reasonable to observe sharp peaks in the nonlinear response of Fig.\ref{2FT_abelain}-(c,d). However, by digressing from the toric code limit (away from the corners), the excitations become dispersed and the streak signals appear as discussed earlier and presented in Fig.\ref{2FT_abelain}-(a, b).

%%%%%%%%%%%%%%%%%%%%%%%%%%%%%%%%%%%%%%%%%%%%%

%%%%%%%%%%%%%%%%%%%%%%%%%%%%%%%%%%%%%%%%%%%%%%

\section{Discussions and Conclusions}\label{conclusion}
We numerically studied the nonlinear response of extended Kitaev model in its Abelian and non-Abelian phases by using two-dimensional coherent spectroscopy. The numerical computations are restricted to finite systems with periodic boundary conditions with the lattice geometry, which has been introduced in Ref.~[\onlinecite{Pedrocchi_Loss}] in order to explicitly determine  the physical and unphysical states of the Kitaev solution. This is important, because physical quantities must be calculated in the physical subspace.

The nonlinear response of the pure  Kitaev model at the isotropic point\cite{choi1} has diagonal and shifted-diagonal streak signals in the 2D frequency space $ \omega_1 $-$ \omega_2 $, however in the extended Kitaev model, these streak signals are very weak and practically no longer exist, where only sharp spots are seen in the response. The sharp spots  are only due to flux excitations and in-gap bound states. 
Away from the triangular phase boundary in the  non-Abelian phase including the isotropic point of the extended Kitaev model we expect similar sharp spots in the nonlinear response.
Distinct signatures of different flux excitations can be discerned within the nonlinear spectroscopic approach. These new features of flux excitations  can not be observed in the linear response.

In the Abelian phase distinct signatures of fractionalized quasiparticles appear in the nonlinear response. 
For two sets of parameters $J_y=J_z=0.3 $ and $J_y=J_z=0.05 $ with $ J_x=1 $ and $ K=0.2 $, we obtain relatively different nonlinear responses.
In the former case, there are strong streak signals  which are signatures of dynamical Majorana fermions ($ \hat{c}_i $) in addition to their $ \omega_1 $-/$ \omega_2 $-intercept as indications of nondynamical Majorana fermions ($ \hat{b}_i^x, \hat{b}_i^y, \hat{b}_i^z $). While in the later case, where one of the Kitaev exchange couplings is much stronger than the others, the streak signals are very weak and only sharp spots show up. The mentioned sharp spots are signature of an effective behavior in terms of conventional toric code. It looks like a cross over between two different dynamical responses in the Abelian phase.

The general form of our presented results are similar to the nonlinear spectroscopic fingerprints of gapped spin liquids, which have been reported in Ref.~[\onlinecite{Nandkishore_Choi2021}]. The difference stems from the fact that in the EKM there are two types of excitations, flux excitations (similar to \emph{e} and \emph{m} excitations in toric code model) and matter excitations that change energy scales and shift the sharp spots in the responses. If we ignore the matter excitations in our calculations, for instance discarding $\varepsilon_{r/p}^{(2)} $ in Eq.(\ref{R_23_frequence}), we will obtain the same responses as in Ref.~[\onlinecite{Nandkishore_Choi2021}]. 
It has to be stressed that the time-reversal symmetry is broken in EKM
in contrast to  the models considered in Ref.~[\onlinecite{Nandkishore_Choi2021}].

The nonlinear responses  presented in this work may be applicable to Kitaev quantum spin liquid candidates in weak magnetic fields. We did not take into account the effect of finite temperature, disorders, and interactions that could be
relevant to explain experimental results, which are proposed for future works.

\section*{ Acknowledgements}
The authors would like to acknowledge W. Choi, M. Kargarian and A. Vaezi for fruitful comments and discussions.

\newpage
\appendix
\begin{widetext}
	\section{The $  Q_u $ matrix} \label{projection_oper}
	In order to specify the physical and unphysical states by determining the sign of the $ D $ in Eq.(\ref{parity_proj}), we first construct the $ Q_u $ matrix.   This is achieved by using the Bogoliubov transformation $ U $ that diagonalizes the Hamiltonian  in Eq.(\ref{H_complex}) as follows
	\begin{equation}
		U\begin{pmatrix} 
			h&\Delta\\
			-\Delta^{*}& -h^{*}
		\end{pmatrix}U^{\dagger}=\begin{pmatrix} 
			\boldsymbol{\varepsilon}&0\\
			0&-\boldsymbol{\varepsilon}
		\end{pmatrix},\quad
		U\begin{pmatrix}
			\hat{f}\\
			\hat{f}^{\dagger}
		\end{pmatrix}=\begin{pmatrix}
			\hat{a}\\
			\hat{a}^{\dagger}
		\end{pmatrix}\quad\Longrightarrow
		\hat{H}_u=\sum_{n}\varepsilon_n\hat{a}^{\dagger}_n\hat{a}_n-\frac{1}{2}\sum_n\varepsilon_n,
	\end{equation}
	where $\boldsymbol{\varepsilon}  $ is a diagonal matrix with entries  $  \varepsilon_n\geq0$. Because $ \hat{a} $ and $ \hat{a}^{\dagger}  $ are hermitian conjugates of each other, the matrix $ U $ can be generally written  as
	\begin{equation}
	\label{bogolibov}
		U=\begin{pmatrix}
			X^*&Y^*\\
			Y&X
		\end{pmatrix}.
	\end{equation}
	The matrix $ U $ can be derived  from eigenvectors of the Hamiltonian. Suppose that $ V^n $ is an eigenvector of the Hamiltonian  with eigenvalue $ \varepsilon_n $,
	\begin{equation}
		V^n=\begin{pmatrix}
			x^n\\y^n
		\end{pmatrix},
	\end{equation}
	where $ x^n$ and $ y^n $ are $N$-dimensional column  vectors. Due to the particle-hole symmetry: $ \Xi\hat{H}_u\Xi^{-1}=-\hat{H}_u $, $ W_n=\Xi V_n $ is also an eigenvector for the Hamiltonian  with the eigenvalue $ -\varepsilon_n $,
	\begin{equation}
		W^n=\Xi V^n=(\tau_x\mathcal{K})V^n=\begin{pmatrix}
			0&1\\1&0
		\end{pmatrix}{V^n}^*=
		\begin{pmatrix}
			{y^n}^*\\{x^n}^*
		\end{pmatrix}.
	\end{equation}
	Using the fact that the eigenvectors of $ \hat{H}_u $ are the column vectors of $ U^{-1} $, we have
	\begin{equation}
		U^{-1}=
		\begin{pmatrix}
			x^1&x^2&\dots&x^N&{y^1}^*&{y^2}^*&\dots&{y^N}^*\\
			y^1&y^2&\dots&y^N&{x^1}^*&{x^2}^*&\dots&{x^N}^*
		\end{pmatrix}.
	\end{equation}
	Since $  U  $ is a unitary matrix, the matrices $ X $ and $ Y $ in (\ref{bogolibov}) are given as follows:
	\begin{equation}
		X=\begin{pmatrix}
			x^1&x^2&\dots&x^N
		\end{pmatrix}^T,\quad
		Y=\begin{pmatrix}
			y^1&y^2&\dots&y^N
		\end{pmatrix}^T.
	\end{equation}
The physical and unphysical states are determined by specifying  the sign of $\hat{D}$ operator: $ \hat{D}=(-1)^{\theta}\det(Q_u)\hat{\pi}_{\chi}\hat{\pi}_a $, where
	\begin{align}
		&\hat{\pi}_{\chi}=\hat{\pi}_{\chi_x}\hat{\pi}_{\chi_y}\hat{\pi}_{\chi_z},\nonumber\\
		&\hat{\pi}_{\chi_{\alpha}}=\prod_{\mu=0}^{N-1}(1-2\hat{\chi}^{\dagger}_{\mu_\alpha}\hat{\chi}_{\mu_\alpha})=\prod_{\mu=0}^{N-1}i\hat{b}_{\mu A}^{\alpha}\hat{b}_{\mu B}^{\alpha}=\prod_{\mu}\hat{u}_{\mu_\alpha},\nonumber\\
		&\hat{\pi}_{a}=\prod_{k=0}^{N-1}(1-2\hat{a}^{\dagger}_{k}\hat{a}_{k})=(-1)^{\sum_k\hat{a}^{\dagger}_k\hat{a}_k},
	\end{align}
	and $ Q_u $ is an orthogonal transformation defined in Ref.~[\onlinecite{Pedrocchi_Loss}] as given below,
	\begin{equation}
		\label{c_and_normal_mode_Q}
		(\hat{b}_1',\hat{b}_1'',\hat{b}_2',\hat{b}_2'',\dots,\hat{b}_N',\hat{b}_N'')=(\hat{c}_1,\hat{c}_2,\hat{c}_3,\dots,\hat{c}_N)Q_u,
	\end{equation}
	where   $ \hat{b}' $ and $ \hat{b}'' $ are fermion operators that are related  to the  canonical fermion operators according to the following relations
	\begin{equation}
		\label{normal_mode}
		\hat{a}_k=\frac{1}{2}(\hat{b}_k'+i\hat{b}_k''),\quad \hat{a}_k^{\dagger}=\frac{1}{2}(\hat{b}_k'-i\hat{b}_k'')\quad
		\Longrightarrow
		\begin{pmatrix}
			\hat{a}\\
			\hat{a}^{\dagger}
		\end{pmatrix}=\frac{1}{2}
		\begin{pmatrix}
			1&i\\
			1&-i
		\end{pmatrix}
		\begin{pmatrix}
			\hat{b}'\\
			\hat{b}''
		\end{pmatrix}.
	\end{equation}
	To find the $ Q_u $ matrix,  we need to find the transformation between $ \hat{b}' $/$ \hat{b}'' $  
	and $ \hat{c}_A$/$ \hat{c}_B$ fermion operators. Firstly,  we use Eq.(\ref{complex_fermion}) and Eq.(\ref{U_transp}),
	\begin{equation}
		\label{a_intermsof_c}
		\begin{pmatrix}
			\hat{a}\\
			\hat{a}^{\dagger}
		\end{pmatrix}=
		U\begin{pmatrix}
			\hat{f}\\
			\hat{f}^{\dagger}
		\end{pmatrix},\quad
		\begin{pmatrix}
			\hat{f}\\
			\hat{f}^{\dagger}
		\end{pmatrix}=\frac{1}{2}
		\begin{pmatrix}
			1&i\\
			1&-i
		\end{pmatrix}
		\begin{pmatrix}
			\hat{c}_A\\
			\hat{c}_B
		\end{pmatrix}
		\quad\Longrightarrow
		\begin{pmatrix}
			\hat{a}\\
			\hat{a}^{\dagger}
		\end{pmatrix}=\frac{1}{2}U
		\begin{pmatrix}
			1&i\\
			1&-i
		\end{pmatrix}
		\begin{pmatrix}
			\hat{c}_A\\
			\hat{c}_B
		\end{pmatrix}.
	\end{equation}
Then, we take into account  Eq.(\ref{normal_mode}) and Eq.(\ref{a_intermsof_c}), which lead to
	\begin{equation}
		\label{Q_praim}
		\begin{pmatrix}
			\hat{b}'\\\hat{b}''
		\end{pmatrix}=Q_u'
		\begin{pmatrix}
			\hat{c}_A\\\hat{c}_B
		\end{pmatrix},\quad
		Q_u'=
		\begin{pmatrix}
			\Re[A]&\Im[B]\\
			-\Im[A]&\Re[B]\\
		\end{pmatrix},
	\end{equation}
where $ A=X+Y$ and $ B=X-Y $. 
According to the labelling of fermion operators within each unit cell we have
%%%%%%%%%%%%%%%%%

%%%%%%%%%%%%%%%%%%
	\begin{equation}
		\label{R_c}
		\begin{pmatrix}
			\hat{c}_A\\\hat{c}_B
		\end{pmatrix}=
		\begin{pmatrix}
			\hat{c}_2\\\hat{c}_4\\\vdots\\\hat{c}_{2N}\\
			\hat{c}_1\\\hat{c}_3\\\vdots\\\hat{c}_{2N-1}
		\end{pmatrix}=
		R_c
		\begin{pmatrix}
			\hat{c}_1\\\hat{c}_2\\\hat{c}_3\\\hat{c}_4\\\vdots\\\vdots\\\vdots\\\hat{c}_{2N}
		\end{pmatrix},\quad
		R_c=
		\begin{pmatrix}
			0&1&0&0&0&\dots&0&0&0\\
			0&0&0&1&0&\dots&0&0&0\\
			&&&&&\vdots&&&\\
			0&0&0&0&0&\dots&0&0&1\\
			1&0&0&0&0&\dots&0&0&0\\
			0&0&1&0&0&\dots&0&0&0\\
			&&&&&\vdots&&&\\
			0&0&0&0&0&\dots&0&1&0
		\end{pmatrix}_{2N\times 2N},
	\end{equation}

	\begin{equation}
		\label{R_b}
		\begin{pmatrix}
			\hat{b}'\\\hat{b}''
		\end{pmatrix}=
		\begin{pmatrix}
			\hat{b}_1'\\\hat{b}_2'\\\vdots\\\hat{b}_{N}'\\
			\hat{b}_1''\\\hat{b}_2''\\\vdots\\\hat{b}_{N}''\\
		\end{pmatrix}=
		R_b
		\begin{pmatrix}
			\hat{b}_1'\\\hat{b}_1''\\
			\hat{b}_2'\\\hat{b}_2''\\
			\vdots\\\vdots\\\hat{b}_{N}'\\\hat{b}_{N}''
		\end{pmatrix},\quad
		R_b=
		\begin{pmatrix}
			1&0&0&0&0&\dots&0&0&0\\
			0&0&1&0&0&\dots&0&0&0\\
			&&&&&\vdots&&&\\
			0&0&0&0&0&\dots&0&1&0\\
			0&1&0&0&0&\dots&0&0&0\\
			0&0&0&1&0&\dots&0&0&0\\
			&&&&&\vdots&&&\\
			0&0&0&0&0&\dots&0&0&1
		\end{pmatrix}_{2N\times 2N}.
	\end{equation}
	Finally, using Eqs.(\ref{Q_praim}), (\ref{R_c}), and (\ref{R_b}), we arrive at the desired relation,
	\begin{equation}
		(\hat{b}_1',\hat{b}_1'',\dots,\hat{b}_N',\hat{b}_N'')=
		(\hat{c}_1,\hat{c}_2,\hat{c}_3,\dots,\hat{c}_{2N})\bigg[
		R_b^{-1}Q_u'R_c
		\bigg]^T\quad\Longrightarrow
		Q_u=\bigg[
		R_b^{-1}Q_u'R_c
		\bigg]^T.
	\end{equation}
With this important matrix we can find the parity $ \hat{\pi}_c=\prod_{\mu=1}^{N}i\hat{c}_{\mu A} \hat{c}_{\mu B}$ 
in terms of the parity of canonical matter fermions\cite{Pedrocchi_Loss}:
\begin{equation}
	\hat{\pi}_c=\det(Q_u)\hat{\pi}_a.
\end{equation}

%%%%%%%%%%%%%%%%%%%%%%%%%%%%%%
	\section{Gauge configurations, Matrix elements, and $ R^{(l),z}_{zzz} $ correlation functions}\label{mat_element}
	\begin{figure*}[!htb]
		\centering
		\includegraphics[scale=1]{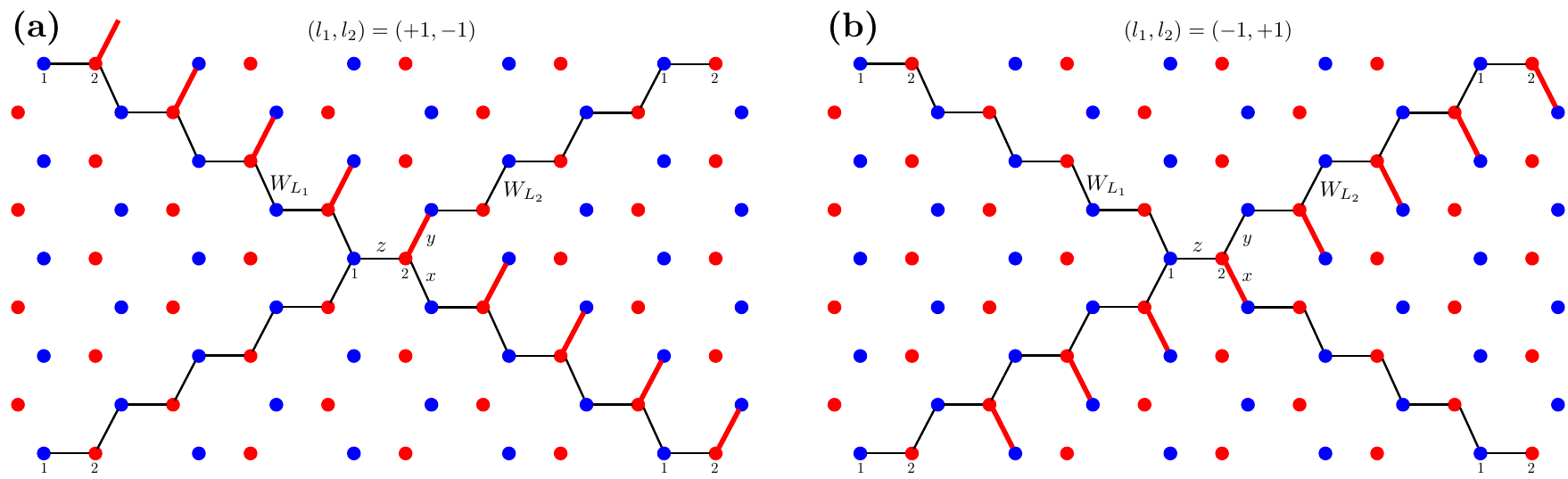}
		\caption{Gauge configuration for two degenerate ground states with different topological labels for a finite system with $L_1=L_2=4 $ and $M=0 $. (a) and (b) show the gauge configuration that we adopted for $ g_{+-}(\hat{\chi}^{\dagger})|\mathcal{G}\rangle $ and  $ g_{-+}(\hat{\chi}^{\dagger})|\mathcal{G}\rangle $, respectively. The black strings links  represent  the topological loop operators $ W_{L_1} $ and $ W_{L_2} $,  which are reduced to the product of $ u_{ij} $'s on these links in a fixed gauge configuration. The red links show the position of links, where $ u_{ij}=-1$.
	}
		\label{gauge_l1_l2}
	\end{figure*}
	
According to subsection (\ref{phy_unphy_states}), the ground state $ |G\rangle $ in the non-Abelian phase  has a 3-fold degeneracy on a torus  with topological labels $ (+1,-1) $, $ (-1,+1) $, and $ (-1,-1) $. Fig.~\ref{gauge_l1_l2}-(a) and Fig.~\ref{gauge_l1_l2}-(b) show the gauge configurations  $ g_{+-}(\hat{\chi}^{\dagger})|\mathcal{G}\rangle $ and $ g_{-+}(\hat{\chi}^{\dagger})|\mathcal{G}\rangle $ for a finite system; designed by red links $ u_{\langle ij\rangle_{\alpha}}=-1 $. The  gauge configuration $ g_{--}(\hat{\chi}^{\dagger})|\mathcal{G}\rangle $ is obtained by taking into account the gauge configurations presented in Figs.~\ref{gauge_l1_l2}-(a, b),  simultaneously, i.e.,  $ g_{--}(\hat{\chi}^{\dagger})=g_{+-}(\hat{\chi}^{\dagger})g_{-+}(\hat{\chi}^{\dagger}) $.

To calculate the matrix elements   in the correlation functions  $ R^{(l),z}_{zzz} $, it is necessary to find the relation between the matter ground states of different flux sectors. Let $ |\mathcal{M}_{F_1}\rangle $ and 
	$ |\mathcal{M}_{F_2}\rangle $ be the ground states of flux sectors $ F_1 $ and $ F_2 $, respectively. Creation and annihilation operators in each sector 	are given as follows,

	\begin{equation}
	\begin{pmatrix}
	\hat{a}_{F_1}\\
	\hat{a}^{\dagger}_{F_1}
	\end{pmatrix}
		=\begin{pmatrix}
			X^*_{F_1}&Y^*_{F_1}\\
			Y_{F_1}&X_{F_1}
		\end{pmatrix}
		\begin{pmatrix}
			\hat{f}\\
			\hat{f}^{\dagger}
		\end{pmatrix}
		\quad,\quad
		\begin{pmatrix}
		\hat{a}_{F_2}\\
		\hat{a}^{\dagger}_{F_2}
		\end{pmatrix}
		=\begin{pmatrix}
			X^*_{F_2}&Y^*_{F_2}\\
			Y_{F_2}&X_{F_2}
		\end{pmatrix}
		\begin{pmatrix}
			\hat{f}\\
			\hat{f}^{\dagger}
		\end{pmatrix},
	\end{equation}
	which are related to each other by the following transformation:
	\begin{equation}
		\label{relate_flux_to_flux}
		\begin{pmatrix}
			\hat{a}_{F_2}\\
			\hat{a}^{\dagger}_{F_2}
		\end{pmatrix}=
		\begin{pmatrix}
			\mathcal{X}^*_{F_2,F_1}&\mathcal{Y}^*_{F_2,F_1}\\
			\mathcal{Y}_{F_2,F_1}&\mathcal{X}_{F_2,F_1}
		\end{pmatrix}
		\begin{pmatrix}
			\hat{a}_{F_1}\\
			\hat{a}^{\dagger}_{F_1}
		\end{pmatrix}\quad,\quad
		\begin{pmatrix}
			\mathcal{X}^*_{F_2,F_1}&\mathcal{Y}^*_{F_2,F_1}\\
			\mathcal{Y}_{F_2,F_1}&\mathcal{X}_{F_2,F_1}
		\end{pmatrix}=
		\begin{pmatrix}
			X^*_{F_2}X^T_{F_1}+Y^*_{F_2}Y^T_{F_1}&X^*_{F_2}Y^{\dagger}_{F_2}+Y^*_{F_2}X^{\dagger}_{F_1}\\
			X_{F_2}Y_{F_1}^T+Y_{F_2}X^T_{F_1}&X_{F_2}X^{\dagger}_{F_1}+Y_{F_2}Y^{\dagger}_{F_1}
		\end{pmatrix}.
	\end{equation}
	According to Refs.~[\onlinecite{Knolle_Chalker_Moessner2015},\onlinecite{Blaizot}], the two ground states
	obey the following relation
	\begin{equation}
	\label{ground_state_in_two_flux_sector}
		|{M_{F_2}}\rangle=\sqrt{|\det(\mathcal{X}_{F_2,F_1})|}e^{-\frac{1}{2}\hat{a}_{F_1}^{\dagger}\mathcal{F}_{F_2,F_1}\hat{a}^{\dagger}_{F_1}}|M_{F_1}\rangle\quad,\quad \mathcal{F}_{F_2,F_1}=\mathcal{X}_{F_2,F_1}^{*-1}\mathcal{Y}_{F_2,F_1}^*.
	\end{equation}
Moreover, we need to write $ \hat{Z}_{\mu} $ in terms of gauge and matter Majorana fermions,
	\begin{equation}
		\hat{Z}_{\mu}=\hat{\sigma}^z_{\mu A}+\hat{\sigma}^z_{\mu B}=i\hat{b}_{\mu A}\hat{c}_{\mu A}+i\hat{b}_{\mu B}\hat{c}_{\mu B}=
		\hat{\chi}_{\mu_z}^{\dagger}(i\hat{c}_{\mu A}+\hat{c}_{\mu B})+\hat{\chi}_{\mu_z}(i\hat{c}_{\mu A}-\hat{c}_{\mu B}).
	\end{equation}
	
We present the details of calculations of the matrix elements that we need to obtain the  response in the Abelian and non-Abelian phase. With the gauge configurations that we adopted for the ground states in these phases, the matrix elements will have the same structures and relations. Therefore, we focus on the states in Eq.(\ref{phy_states_non_abelian}). For simplicity, we ignore the index $ +-  $ on these states.  The matrix element $ \langle P_{\mu}|\hat{Z}_{\mu}|G\rangle $   in the matter sector  is reduced as the following,
	\begin{align}
		\langle P_{\mu}|\hat{Z}_{\mu}|G\rangle=
		\langle\mathcal{M}_{\mu}|\overline{\hat{a}}_r(i\hat{c}_{\mu A}+\hat{c}_{\mu B})|\mathcal{M}\rangle.
	\end{align}
%%%%%%%%%%%%%%%%%%%%%%%
By using  Eqs.(\ref{a_intermsof_c}) and (\ref{relate_flux_to_flux}), we write these operators in terms of the canonical matter fermions  in the 0-flux sector,
	\begin{align}
		&i\hat{c}_{\mu A}+\hat{c}_{\mu B}=2i\bigg[
		Y^T_{\mu s}\hat{a}_s+X^{\dagger}_{\mu s}\hat{a}_s^{\dagger}
		\bigg],\nonumber\\
		&\overline{\hat{a}}_r=\mathcal{X}^*_{\mu,rr'}\hat{a}_{r'}+
		\mathcal{Y}_{\mu,rr'}^*\hat{a}^{\dagger}_{r'}.
	\end{align}
	Therefore,
	\begin{align}
		\langle P_{\mu}|\hat{Z}_{\mu}|G\rangle=&
		2i\sqrt{|\det(\mathcal{X}_{\mu})|}X_{\mu s}^{\dagger}\ 
		\langle\mathcal{M}|\bigg[
		1-\frac{1}{2}\hat{a}_n\mathcal{F}_{\mu,nm}^{\dagger}\hat{a}_m+\dots
		\bigg]
		\bigg[
		\mathcal{X}^*_{\mu,rr'}\hat{a}_{r'}+
		\mathcal{Y}_{\mu,rr'}^*\hat{a}^{\dagger}_{r'}
		\bigg]
		|\mathcal{M}\rangle,\nonumber\\
		&=
		2i\sqrt{|\det(\mathcal{X}_{\mu})|}X_{\mu s}^{\dagger}\ 
		\bigg[
		\mathcal{X}^{\dagger}_{\mu}+
		\frac{1}{2}
		(\mathcal{F}^*_{\mu}\mathcal{Y}^{\dagger}_{\mu}-\mathcal{F}^{\dagger}_{\mu}\mathcal{Y}^{\dagger}_{\mu})
		\bigg]_{sr},
	\end{align}
and according to Eqs.(\ref{relate_flux_to_flux}) and (\ref{ground_state_in_two_flux_sector}), $ \mathcal{F}_{\mu}=\mathcal{X}^{*-1}_{\mu} \mathcal{Y}^{*}_{\mu}$, $ \mathcal{X}_{\mu}=X_{\mu}X^{\dagger}+Y_{\mu}Y^{\dagger} $, and   $ \mathcal{Y}_{\mu}=X_{\mu}Y^{T}+Y_{\mu}X^{T} $ . With the implementation  of the identity: 
	$  \mathcal{F}^*_{\mu}\mathcal{Y}^{\dagger}_{\mu}-\mathcal{F}^{\dagger}_{\mu}\mathcal{Y}^{\dagger}_{\mu}=
	2(\mathcal{X}^{-1}_{\mu}-\mathcal{X}^{\dagger}_{\mu})
	$, 
	we arrive at the simple relation:
	\begin{equation}
		\langle P_{\mu}|\hat{Z}_{\mu}|G\rangle=
		2i\sqrt{|\det(\mathcal{X}_{\mu})|}\ 
		\bigg[
		X^{\dagger}\mathcal{X}_{\mu}^{-1}
		\bigg]_{\mu r}.
	\end{equation}
By performing similar steps, we get
	\begin{equation}
		\langle R_{\nu}|\hat{Z}_{\mu}|Q_{\mu\nu}\rangle=
		2i\sqrt{|\det(\mathcal{X}_{\nu,\mu\nu})|}\ 
		\bigg[
		Y^{\dagger}_{\mu\nu}
		\mathcal{X}^{-1}_{\nu,\mu\nu}
		\bigg]_{\mu r},\quad\text{where}\quad \mathcal{X}_{\nu,\mu\nu}=X_{\nu}X^{\dagger}_{\mu\nu}+Y_{\nu}Y_{\mu\nu}^{\dagger}.
	\end{equation}

The existence of translational invariance in the zero-flux state $ |G\rangle $ in the Abelian and non-Abelian phases allows us to replace 
$\sum_{\mu\nu}$ by $N\sum_{\mu\neq 1}\delta_{\nu,1}$ in  Eq.(\ref{phy_contributions}). Finally, as an example, one can arrive at  the following summation for $  R^{(1,2),z}_{zzz}(\tau_2,\tau_1,0) $,
	\begin{align}
	\label{phy_equ_R12}
	R^{(1,2),z}_{zzz}(\tau_2,\tau_1,0)=N\sum_{\nu\neq 1}
	e^{i[E({1\nu})-E_2]\tau_2}\bigg\{
	&\sum_{p=1}^{N} e^{i\varepsilon_p^{(1)})\tau_1}
	\langle G|\hat{Z}_{1}
	|{P_{1}}\rangle
	\langle{P_{1}}|\hat{Z}_{\nu}
	|{Q_{1\nu}}\rangle\nonumber\\
	\times&\sum_{r=1}^{N}
	e^{-i\varepsilon_r^{(1)}(\tau_1+\tau_2)}
	\langle{Q_{1\nu}}|\hat{Z}_{\nu}
	|{R_{1}}\rangle
	\langle{R_{1}}|\hat{Z}_{1}
	|G\rangle\nonumber\\
	+
	&\sum_{p=1}^{N} e^{i\varepsilon_r^{(\nu)}\tau_1}
	\langle G|\hat{Z}_{\nu}
	|{P_{\nu}}\rangle
	\langle{P_{\nu}}|\hat{Z}_{1}
	|{Q_{1\nu}}\rangle\nonumber\\
	\times&\sum_{r=1}^{N}
	e^{-i\varepsilon_r^{(1)}(\tau_1+\tau_2)}
	\langle{Q_{1\nu}}|\hat{Z}_{\nu}
	|{R_{1}}\rangle
	\langle{R_{1}}|\hat{Z}_{1}
	|G \rangle\bigg\},
\end{align}
where $ E(1\nu) $ and $ E_2 $ are value of Eq.(\ref{gse}) for  the state $ |Q_{1\nu}\rangle $ and 2-flux sate $ |P_1\rangle, |P_{\nu}\rangle, |R_1\rangle$.

	\begin{figure*}[!t]
	\center
	\includegraphics[scale=1.2]{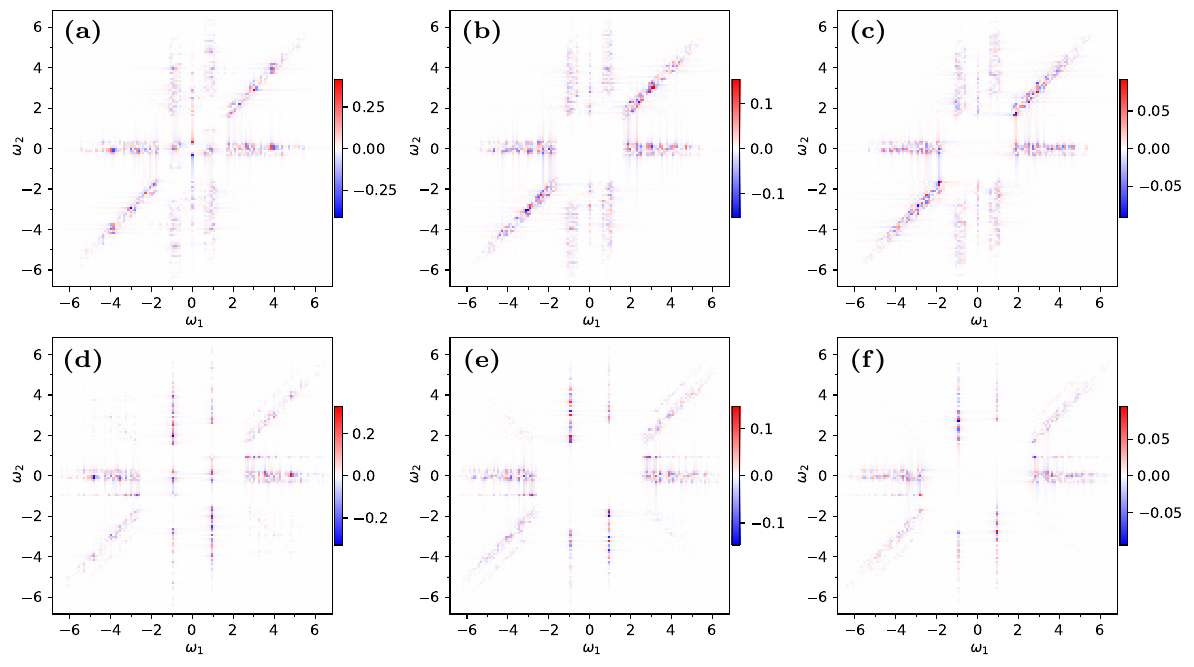}
	\caption{The difference between susceptibilities at two different finite sizes of system at the isotropic point with $ K=0.2 $.  The top row shows the results $ \chi^{(3),z}_{zzz}(\omega_2,\omega_1,0; L_2) -  \chi^{(3),z}_{zzz}(\omega_2,\omega_1,0; L_1)$,
		for (a) $L_2=15, L_1=7$,  (b) $L_2=27, L_1=15$ and (c) $L_2=33, L_1=27$. As the size of system is increased the difference of susceptibility at two successive sizes is reduced strongly. The bottom row exhibits similar results for
		$ \chi^{(3),z}_{zzz}(\omega_2,0,\omega_1; L_2) -  \chi^{(3),z}_{zzz}(\omega_2,0,\omega_1; L_1)  $, where $L_2$ and $L_1$ are respectively (d) $15, 7$, (e) $27, 15$ and (f) $33, 27$.
		Some streak signals can be seen in the above plots. Although these signals are present in the response of Fig.~\ref{2FT_phy} they are so weak that are revealed only by subtracting the response at two successive system sizes.
	} 
	\label{finite_size}
\end{figure*}

\section{Finite size effects}\label{finte_size_effects}
Here, we show that the finite size effects in the nonlinear responses are weak and a system size of $33\times33$ unit cells ($2178$ spins) represent a reasonable result for the nonlinear susceptibilities in the thermodynamic limit.
The top row of Fig.~\ref{finite_size} show the difference value of the normalized susceptibility $ \chi^{(3),z}_{zzz}(\omega_2,\omega_1,0; L) $ for two different sizes.  Fig.~\ref{finite_size}-(a) represents the difference 
$ \chi^{(3),z}_{zzz}(\omega_2,\omega_1,0; L=15) -  \chi^{(3),z}_{zzz}(\omega_2,\omega_1,0; L=7)$ and similar
results have been plotted in (b) for $L=27, 15$ and (c) for $L=33, 27$. The color bar shows a decrease in the absolute value as the size increases, which justifies our claim. Similar values have been plotted in the bottom row of Fig.~\ref{finite_size} for $ \chi^{(3),z}_{zzz}(\omega_2,0,\omega_1; L) $, where (d) shows the results for $L=15, 7$, (e) for  $L=27, 15$ and (f) for $L=33, 27$. 
The weak finite size effect is expected since the underlying system is gapped.

\section{Response function versus exchange coupling \label{response-J}}
In order to further confirm our statement about the presence/absence of the streak signals 
in the Abelian phase of our model (cf.  the last part of Sec. \ref{nonlinaer_result}), we have plotted the nonlinear susceptibility in Fig.~\ref{response_evolution} for several values of $J$ along the white path shown in Fig.~\ref{RZG}-(b).  In these plots, $J=\frac{J_y}{J_x}=\frac{J_z}{J_x}$,  and $K=0.2$, the top panel shows  $ \chi^{(3),z}_{zzz}(\omega_2,\omega_1,0) $ and the bottom panel represents  $\chi^{(3),z}_{zzz}(\omega_2,0,\omega_1)$.It has to be mentioned that the plots in Fig.~\ref{response_evolution} represent unnormalized data that are indicated by their corresponding color bar, which shows the evolution of the response function with respect to $J$. The plots for $J=0.05$ show sharp spots revealing the contribution from a single excitation, while the plots for $J=0.25, 0.45$ represent streak signals of several excitation modes.  For the finite size of the underlying lattice $ L_1=L_2=28 $, the plots of $J=0.65$ show the crossover from the streak signals to sharp peaks of the flux excitations in non-Abelian phase, where the latter become obvious for $J=0.85$.

\begin{figure*}[!t]
	\center
	\includegraphics[scale=1.5]{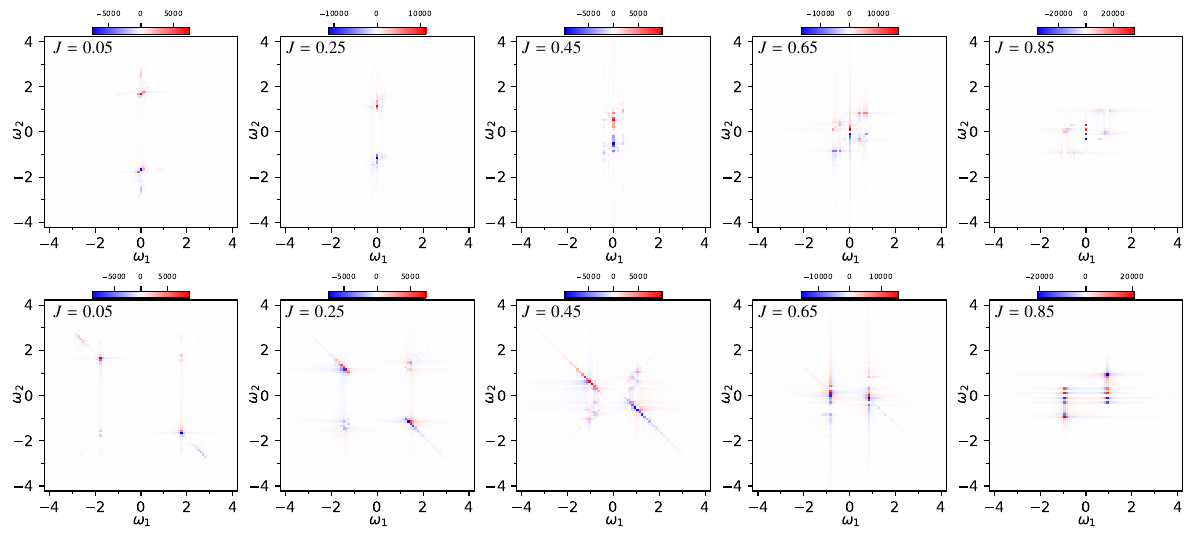}
	\caption{The top and bottom panels show susceptibilities $ \chi^{(3),z}_{zzz}(\omega_2,\omega_1,0) $ and $ \chi^{(3),z}_{zzz}(\omega_2,0,\omega_1) $, respectively. They are
unnormalized 2D nonlinear response for several values of $J=0.05, 0.25, 0.45, 0.65, 0.85$ along the white path of Fig.~\ref{RZG}-(b). 
In all plots the lattice geometry is $ L_1=L_2=28 $ and $ M=0 $.  
	} 
	\label{response_evolution}
\end{figure*}

\section{Effective Hamiltonian}\label{Effective_Hamiltonian}

Let us consider an extreme limit and assume $ J_x=J_y=K=0 $ in the EKM. For  $ J_z>0 $, the spin configurations in which two spins on the z-link are aligned up or down ($ |\uparrow\uparrow\rangle $ or $ |\downarrow\downarrow\rangle $), form the degenerate ground state subspace with the energy $ E_0=-NJ_z $. 
The two dimensional ground state subspace (on each z-link) can be considered as an effective spin $ |\Uparrow\rangle $ or $ |\Downarrow\rangle $ on the lattice, which is shown in Fig.~\ref{effective_spin}-(b). The effective Pauli matrices act on the effective spins as  sketched in Fig.~\ref{effective_spin}-(c).  
For nonzero values of $J_x, J_y$ and  $ K $, and in a perturbative regime where $J_x, J_y, K \ll J_z$, 
the Hamiltonian of the system is written as $ H=H_0+V $,
\begin{equation}
H_0=-J_z\sum_{\text{i}\in\text{ z-links}}T^z_i,\quad\quad V=-J_x\sum_{\text{i}\in\text{ x-links}}T^x_i
-J_y\sum_{\text{i}\in\text{ y-links}}T_i^y
-K\sum_{\substack{\langle ik\rangle_{\alpha},\langle kj\rangle_{\beta}\\\gamma\perp\alpha,\beta}}\hat{\sigma}^{\alpha}_i\hat{\sigma}^{\gamma}_k\hat{\sigma}^{\beta}_j,
\end{equation}
where $ T^{\alpha}_i=\hat{ \sigma}^{\alpha}_k\hat{\sigma}^{\alpha}_j  $. 
The three-spin interactions can also be written in terms of $ T_i^{\alpha}$'s. For instance, consider the plaquette $ p_1 $ in Fig.~\ref{effective_spin}-(a), where all possible three-spin interactions  are expressed in the following,
\begin{align}
\label{Three_spin_on_p}
-K&(\hat{ \sigma}^x_1\hat{ \sigma}^z_2\hat{ \sigma}^y_3
+\hat{ \sigma}^y_2\hat{ \sigma}^x_3\hat{ \sigma}^z_4
+\hat{ \sigma}^z_3\hat{ \sigma}^y_4\hat{ \sigma}^x_5
+\hat{ \sigma}^x_4\hat{ \sigma}^z_5\hat{ \sigma}^y_6
+\hat{ \sigma}^y_5\hat{ \sigma}^x_6\hat{ \sigma}^z_1
+\hat{ \sigma}^z_6\hat{ \sigma}^y_1\hat{ \sigma}^x_2
)\nonumber\\
&=i K([T_1^x T_2^y]+[T_2^y T_R^z]+[T_R^z T_3^x]+[T_3^x T_4^y]
+[T_4^y T_L^z]+[T_L^z T_1^x]),
\end{align}
where we used the square brackets to indicate  the distinction between three-spin interaction terms and the interactions in the first and second terms of $ V $.  Suppose that $ P_0 $ and $ Q_0 $ are the projection operators onto the  ground state subspace  and excited states of $ H_0$, respectively. By using the Brillouin-Wigner pertrurbaation approach the effective Hamiltonian of the system with energy $ E\approx E_0 $ is,
\begin{equation}
H_{\text{eff}}=P_0H_0P_0+P_0VP_0+P_0VG_0'VP_0+P_0VG_0'VG_0'VP_0+\dots
\end{equation}
where $ G_0'=\frac{1}{E_0-H_0}Q_0$ is the Green's function.

	\begin{figure*}[!t]
	\center
	\includegraphics[scale=1.1]{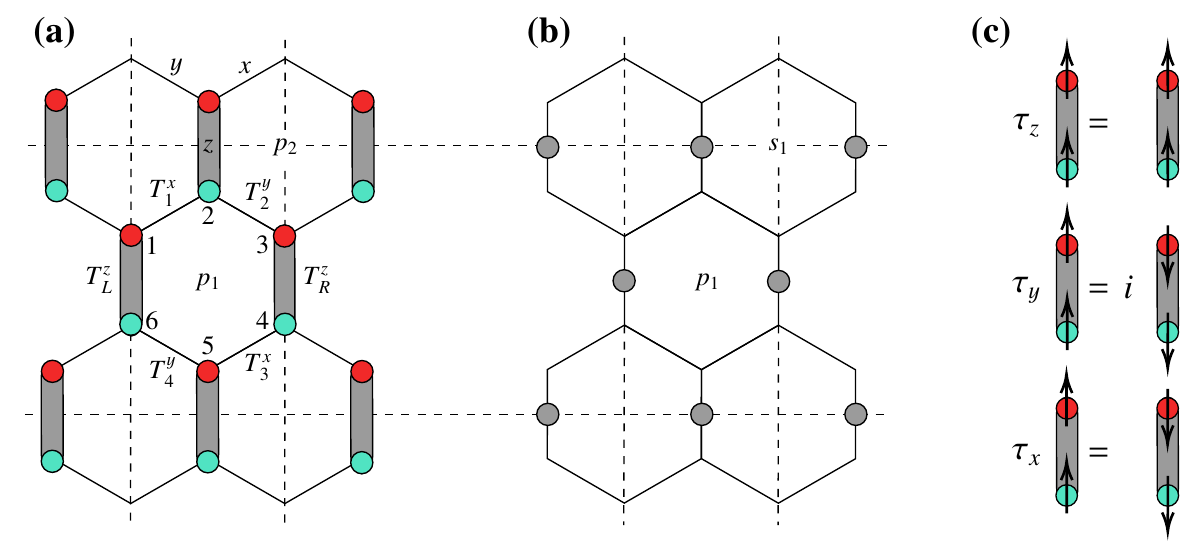}
	\caption{The strong links of EKM form an effective lattice in the large $ J_z $ limit ($J_z \gg J_x, J_y, K$). (a) The large z-links are shown graphically with thick gray bonds and $T_i^{\alpha}$'s are Kitaev exchange interactions. (b) The gray circles represent the  effective spins on the effective square lattice constructed with the dashed lines. (c) The act of effective Pauli operators on an effective spin is shown schematically.
	    } 
	\label{effective_spin}
\end{figure*}

The $n$-th order of  perturbation term ($H_{\text{eff}}^{(n)}$) along with some of the interaction terms are given in the following, 
\begin{align}
&H_{\text{eff}}^{(1)}=0\\
&H_{\text{eff}}^{(2)}=\text{const.}-\frac{K^2}{2J_z}\sum_{i}Q_{p_i} ;\quad  [T_1^x T_2^y][T_3^x T_4^y]\; \mbox{and} \;   [T_3^x T_4^y][T_1^x T_2^y] ,\\
& H_{\text{eff}}^{(3)}=\text{const.}-\left(\frac{J_x K^2}{4J_z^2}+\frac{J_yK^2}{4J_z^2}\right)\sum_{i}Q_{p_i} ; \quad [T_3^x T_4^y][T_2^y T_R^z]T_1^x , \cdots, \\
&
 H_{\text{eff}}^{(4)} =\text{const.}-\left(\frac{K^4+J_x^2K^2+J_y^2K^2+J_x^2J_y^2}{16J^3_z}\right)\sum_{i}Q_{p_i};\quad
 [T_4^y T_L^z][T_R^z T_3^x][T_2^y T_R^z][T_L^z T_1^x], [T_4^y T_L^z][T_2^y T_R^z]T_3^x T_1^x ,
T_4^y T_3^x T_2^y T_1^x , \dots,
\end{align}
where for example $ Q_{p_1}=\hat{ \sigma}_2^z\hat{ \sigma}_5^z(\hat{ \sigma}_1^x\hat{ \sigma}_6^y)(\hat{ \sigma}_3^y\hat{ \sigma}_4^x) $ will be equal to $ \hat{\tau}_{\text{up}}^z\hat{\tau}_{\text{down}}^z\hat{\tau}_{\text{left}}^y\hat{\tau}_{\text{right}}^y $  in terms of the effective Pauli matrices on the plaquette $ p_1 $ as shown in Fig.~\ref{effective_spin}-(b)\cite{Kitaev_2006}.  Unlike the pure KM, the first nonzero term in the perturbation expansion appears at the second order since the three-spin interactions of the EKM are made up of two Kitaev interactions (\ref{Three_spin_on_p}). With an appropriate unitary transformation, $ H_{\text{eff}} $ in terms of $ Q_{pi} $'s is transformed into the Kitaev toric  code\cite{Kitaev_2006}.

\end{widetext}

%%%%%%%%%%%%%%%%%%%%%%%%%%%%%%%%%%%%%%%%%%%%%%%%%%%%%%%%%%
%%%%%%%%%%%%%%%%%%%%%%%%%%%%%%%%%%%%%%%%%%%%%%%%%%%%%%%%%%
%merlin.mbs apsrev4-1.bst 2010-07-25 4.21a (PWD, AO, DPC) hacked
%Control: key (0)
%Control: author (8) initials jnrlst
%Control: editor formatted (1) identically to author
%Control: production of article title (-1) disabled
%Control: page (0) single
%Control: year (1) truncated
%Control: production of eprint (0) enabled
%

%%%%%%%%%%%%%%%%%%%%%%%%%%%%%%%%%%%%%%%%%%%%%%%%%%%%%%%%%%
%%%%%%%%%%%%%%%%%%%%%%%%%%%%%%%%%%%%%%%%%%%%%%%%%%%%%%%%%%

%\bibliography{mybib1}

\begin{thebibliography}{75}
\bibitem [{com()}]{comment3}%
  \BibitemOpen
  \href@noop {} {}\bibinfo {note} {We computed fast fourier transformation with
  Python package for calculation two-dimensional discrete fast fourier
  transformation.}\BibitemShut {Stop}%
\end{thebibliography}

\end{document}